\documentclass[prb,twocolumn,showpacs,amsmath,amssymb]{revtex4}

\usepackage{graphicx}
\usepackage{dcolumn}
\usepackage{bm}
\usepackage{epsfig}

\begin{document}

\title{{\em I-V} curve signatures of nonequilibrium-driven band gap collapse in magnetically ordered zigzag graphene nanoribbon two-terminal devices}

\author{Denis A. Areshkin}
\affiliation{Department of Physics and Astronomy, University
of Delaware, Newark, DE 19716-2570, USA}
\author{Branislav K. Nikoli\' c}
\affiliation{Department of Physics and Astronomy, University
of Delaware, Newark, DE 19716-2570, USA}

\begin{abstract}
Motivated by the very recent fabrication of sub-10-nm-wide semiconducting graphene nanoribbons [X. Li {\em et al.}, Science {\bf 319}, 1229 (2008)], where
some of their band gaps extracted from transport measurements were closely fitted to density functional theory predictions for magnetic ordering along zigzag edges
that is responsible for the insulating ground state, we compute current-voltage (\mbox{\em I-V}) characteristics  of finite-length zigzag graphene
nanoribbons (ZGNR) attached to metallic contacts. The transport properties of such devices, at source-drain bias voltages beyond the linear response
regime, are obtained using the nonequilibrium Green function formalism combined with the mean-field version of the Hubbard model fitted to reproduce
the local spin density approximation description of magnetic ordering. Our results indicate that magnetic ordering and the corresponding  band gap in
ZGNR can be {\em completely eliminated} by passing large enough DC current through it. The threshold voltage for the onset of band gap collapse depends on the
ZGNR length and the contact transparency. If the contact resistance is adjusted to experimentally measured value of $\simeq 60$~k$\Omega$, the threshold voltage
for sub-10-nm-wide ZGNR with inter-contact distance of $\simeq 7$ nm is $\approx 0.4$ V.  For some device setups, including 60~k$\Omega$ contacts, the room
temperature \mbox{\em I-V} curves demonstrate step-like current increase by an order of magnitude  at the threshold voltage, and can exhibit a hysteretic behavior
as well. On the other hand, poorly transmitting contacts can completely eliminate abrupt jump in the \mbox{\em I-V} characteristics. The threshold voltage increases with the ZGNR length (e.g., reaching $\approx 0.8$ V for $\simeq 13$ nm long ZGNR) which provides possible explanation of why the recent experiments [Wang {\em et al.}, Phys. Rev. Lett. {\bf 100}, 206803 (2008)] on $\sim 100$ nm long GNR field-effect transistors with bias voltage $< 1$ V did not detect the \mbox{\em I-V} curve signatures of the band gap collapse. Thus, observation of predicted abrupt jump in the \mbox{\em I-V} curve of two-terminal devices with short ZGNR channel and transparent metallic contacts will confirm its zigzag edge magnetic ordering via {\em all-electrical} measurements, as well as a current-flow-driven magnetic-insulator--nonmagnetic-metal nonequilibrium phase transition.
\end{abstract}


\maketitle

\section{Introduction} \label{sec:intro}

The recent surprising discovery of graphene~\cite{Geim2007}---a one-atom-thick layer of graphite---has introduced in a short period of
time a plethora of new concepts in condensed matter physics and nanotechnology, despite apparent simplicity of the two-dimensional
honeycomb lattice of carbon atoms that underlies much of its unusual physics revolving around Dirac-like low-energy electronic excitations.~\cite{Neto2009}
Examples include anomalous versions of mesoscopic transport effects,~\cite{Geim2007} topological insulators,~\cite{Kane2005a} and low-dimensional
carbon-based magnetism,~\cite{Son2006,Fernandez-Rossier2007,Wang2008} to name just a few. Since driving a system out of equilibrium
typically corrupts its quantum coherence and suppresses quantum interference effects, basic research experiments have mostly been focused on
the linear response regime.~\cite{Geim2007}

At the same time, vigorous pursuit of carbon nanoelectronics,~\cite{Avouris2007,Avouris2009} envisioned around gated planar graphene
structures that promise to overcome some of the difficulties~\cite{Avouris2007} encountered by carbon nanotubes, has led to increasing number of
experimentally demonstrated top-gated graphene field-effect transistor (FET) concepts. In these setups, micron-size graphene sheets~\cite{Novoselov2004,Meric2008,Lin2009}
or sub-10-nm-wide graphene nanoribbons~\cite{Wang2008a} were employed to demonstrate room temperature graphene-FET operation with ON/OFF current ratios~\cite{Wang2008a}
up to $10^6$, high carrier mobility in sheet-based FETs,~\cite{Meric2008} large critical current densities,~\cite{Novoselov2004} and operating frequency reaching~\cite{Lin2009} $\simeq 26$ GHz.

These experiments pose a challenge for theoretical and computational modeling since they drive graphene nanostructures into far-from-equilibrium regime due to the finite applied bias voltage. The task is more demanding than typical linear response-based analysis~\cite{Cheianov2006,Rycerz2007}  of potential graphene devices due to the need to compute self-consistently developed potential and charge redistribution within the system in nonequilibrium current carrying state in order to keep the gauge invariance~\cite{Christen1996} of the \mbox{\em I-V} characteristics intact.

In addition, the description of experimental devices often requires to include much greater microscopic details~\cite{Reich2002,Gruneis2008}
than captured by simplified effective models that resemble relativistic Dirac Hamiltonian for massless fermions~\cite{Cheianov2006,Tworzydlo2008} or its parent single $\pi$-orbital nearest-neighbor tight-binding Hamiltonian.~\cite{Rycerz2007} These include atomic (such as the presence of hydrogen atoms which passivate edge carbon atoms~\cite{Pisani2007}) and electronic structure (probed by the bias voltage defined energy window around the Dirac point), self-consistent charge transfer effects~\cite{Ohno2000a} that depend on the environment of an atom (tight-binding models are blind with respect to the charge of the system), and possibly more intricate manifestations~\cite{Yang2007c,Zarea2008} of electron-electron interactions in quasi-one-dimensional graphene nanostructures.

For example, unlike the sheets of bulk graphene which can be viewed as a zero-gap semiconductor,~\cite{Meric2008,Lin2009} measured ratios of currents in ON and OFF states $I_{\rm ON}/I_{\rm OFF}$ for room-temperature GNRFET~\cite{Li2008,Wang2008a} reveal that {\em all} of the sub-10-nm-wide nanoribbons underlying the device were {\em semiconducting}. Furthermore, some of the energy gaps extracted from the operation of GNRFETs were closely fitted by the density functional theory (DFT) predictions for magnetic insulating ground state of ZGNR whose band gap is inversely proportional to GNR width. This is in contrast to non-interacting continuum Dirac~\cite{Brey2006} or tight-binding~\cite{Rycerz2007} models of ZGNR which find only metallic nanoribbons with no energy gap around the Fermi level. Even larger band gaps, predicted~\cite{Yang2007c} by more complicated (non-self-consistent) many-body GW treatment of putatively enhanced electron-electron interactions in very narrow ZGNRs, were not seen in these measurements.

A GNR is created by cutting a graphene sheet along two parallel lines. The recently developed nanofabrication techniques for sub-10-nm-wide GNR include direct STM tip drawing~\cite{Tapaszto2008} and chemical derivation.~\cite{Li2008,Wang2008a}  The GNRs produced by the latter technique were used for \mbox{\em I-V} curve measurements in Refs.~\onlinecite{Wang2008a} and \onlinecite{Li2008}.  Their crystallographic orientations were not identified.  However, the fact that the number of $sp^{2}$ bonds per unit length to be cut by chemical derivation in ZGNR is less than the number of bonds in armchair GNR (AGNR) suggests that chemical derivation is more likely to produce ZGNR rather than AGNR.

Both AGNR and ZGNR are predicted to be semiconducting,~\cite{Son2006a} where the origin of their band gap is different. The band gap in AGNR is the consequence of quantum confinement and increased hopping integral between the $\pi$-orbitals on the atoms around the armchair edge caused by slight changes in atomic bonding length.~\cite{Son2006a} On the other hand, the band gap in ZGNR is due to staggered sublattice potential arising due to non-zero spin polarization around the zigzag edges.~\cite{Son2006a}

Although not confirmed by direct probing (such as via sophisticated spin-polarized STM techniques able to detect magnetic moment of individual atoms~\cite{Meier2008}), the possibility of peculiar carbon-based {\em s-p} magnetism (in contrast to conventional magnetism originating from $d$ or $f$ electrons~\cite{Skomski2008}) has been known since the early studies~\cite{Fujita1996} of edge localized states due to special topology~\cite{Ryu2002} of zigzag edges. These states have partially flat (within one-third of the 1D Brillouin zone) subband, thereby generating large peak in the density of states at the Dirac point (i.e., the Fermi energy of undoped graphene). This makes it possible to easily satisfy the Stoner criterion~\cite{Skomski2008} for magnetic ordering when (even tiny~\cite{Fujita1996}) Coulomb interaction is taken into account, which is the most likely~\cite{Pisani2007} way to resolve the instability brought about by the high density of states
at the Fermi level in the nonmagnetic ZGNRs. Furthermore, the study of ZGNR magnetism has recently emerged as one of the major topics of theoretical research on graphene, reignited in part by the DFT calculations within the local spin density approximation (LSDA) that have described properties of such ordering from first principles.~\cite{Son2006,Son2006a,Pisani2007}

In this {\em equilibrium} picture, the ground state electronic configuration of both infinite~\cite{Son2006,Son2006a,Pisani2007} and finite-length~\cite{Dragomirova2008a} ZGNR is characterized by ferromagnetic ordering of spins at each zigzag edge, antiparallel spin orientation between the two edges, and antiferromagnetic coupling between the two edges. Such compensated ferrimagnetic ordering within ZGNR free of defects has zero total magnetic moment. Since opposite spin states occupy different triangular sublattices of the honeycomb lattice, the corresponding staggered potential induces~\cite{Kane2005a} the energy gap. The gap is inversely proportional to the width of the ribbons because the potential  in the middle of the ribbon decreases as the width increases (the band gap vanishes within the room-temperature thermal energy window when the width of GNR reaches $\simeq 80$ nm).~\cite{Pisani2007,Son2006a}

These findings  have  also motivated numerous proposals for applications of ZGNR and graphene nanoislands with zigzag edges in spintronics,~\cite{Son2006,Wimmer2008,Wang2008,Kim2008,Ezawa2008} despite the fact that no true long-range ordering in one-dimension is expected at finite temperatures (for example, at room temperature  the range of magnetic ordering along the edge is quantified by the spin correlation length estimated~\cite{Yazyev2008} to be $\sim 1$ nm). Moreover, virtually all known manifestations of edge magnetic ordering in ZGNR have been predicted within the framework of equilibrium theories~\cite{Son2006,Son2006a,Pisani2007} or linear response transport calculations~\cite{Wimmer2008} which assume vanishingly small bias voltage. Except for the study of its modification, and ultimately destruction, in idealized infinite ZGNR due to the passage of finite ballistic current,~\cite{Gunlycke2007} {\em very little is known} on how such magnetism will manifest in the transport properties of realistic devices where finite-length ZGNR is attached to metallic contacts~\cite{Dragomirova2008a,Kim2008} and biased by finite voltage applied between electrodes, as is the case of experiments on GNRFETs reported in Refs.~\onlinecite{Wang2008a} and ~\onlinecite{Li2008}.

Here we describe the fate of the band gap in two-terminal ZGNR devices  where finite bias voltage brings them into a {\em nonequilibrium} steady transport state. Our results predict that passing a large enough current along ZGNR results in the destruction of spin-polarization around zigzag edges.  This, in turn, causes the collapse of magnetic-ordering-induced band gap, and hence can lead to an {\em abrupt step} in the \mbox{\em I-V} characteristics of ZGNR. Nevertheless, this fundamentally nonequilibrium effect was not observed in recent experimental measurements~\cite{Wang2008a} of the GNRFET \mbox{\em I-V} characteristics. Therefore, the second principal goal of our study is to provide explicit prerequisites for the experimental observation of current-flow-driven collapse of spin-polarized state in ZGNRs and the corresponding magnetic-insulator--nonmagnetic-metal nonequilibrium phase transition.  We note that phenomenologically similar voltage-driven HOMO-LUMO gap collapse in a molecule attached to two electrodes was predicted~\cite{Thygesen2008} when these two levels (broadened by quasiparticle scattering)  hit the bias window simultaneously, which together with our findings emphasize possibility of highly intricate phenomena due to the complexity of nonequilibrium steady state in the finite-bias transport regime.

The paper is organized as follows. Section \ref{sec:Methods} introduces the Hubbard model in the mean-field approximation as a two-parameter fit to {\em ab initio} LSDA. In Sec.~\ref{sec:NewtonRaphson} we describe the Newton-Raphson method used to accelerate the convergence of self-consistent spin-resolved electron density in the nonequilibrium state.  The minimal basis set and the Hubbard model make the relatively expensive Newton-Raphson method much simpler and very efficient when combining with the nonequilibrium Green function (NEGF) techniques.   Section~\ref{sec:Thermodynamics} considers the influence of finite temperature and edge disorder on the spin-polarization of ZGNR in equilibrium. In Sec.~\ref{sec:NonEquilibriumCase}, we present our principal results: (i) the threshold voltage required to destroy the edge spin polarization increases with the ZGNR length, reaching $\approx 0.4$~V and $\approx 0.8$~V for ZGNRs of length $\simeq 7$~nm and $\simeq 13$~nm, respectively (Figs.~\ref{fig:IVMediumSizeStrip} and \ref{fig:IVLongStrip}); (ii) since larger threshold voltages and higher turn-on current may destroy ZGNR, we propose that the length of ZGNR intended for the band gap collapse measurements and, therefore, \mbox{\em I-V} curve probing of the underlying magnetic ordering,  should be of the order of \mbox{$\sim  10$~nm}. We also discuss in Sec.~\ref{sec:NonEquilibriumCase} the influence of the contact quality on the observability of predicted features of the \mbox{\em I-V} characteristics of ZGNR sandwiched between two metallic electrodes. We conclude in Sec.~\ref{sec:conclusion}, while providing technical details of the self-consistent electron density calculations in the nonequilibrium state in Appendix~\ref{sec:appendix_a}.

\section{ZGNR Effective minimal-basis-set self-consistent Hamiltonian as a two-parameter fit to LSDA} \label{sec:Methods}

The texture of magnetic ordering in confined graphene nanostructures, with at least few carbon atoms~\cite{Kumazaki2008} forming a zigzag edge, has been described quantitatively either
by using the mean-field approximation of the  Hubbard (MFAH) model with single $\pi$-orbital per site~\cite{Fujita1996,Fernandez-Rossier2007,Kumazaki2008} or DFT within different approximation schemes for its exchange-correlation density functional (such as LSDA,~\cite{Son2006a} GGA,~\cite{Fernandez-Rossier2007} or hybrid B3LYP~\cite{Pisani2007}). Although DFT goes beyond strictly on-site treatment of the electron-electron interaction $U$ and the nearest-neighbor hopping $t$ in the MFAH model, the parameters of the latter~\cite{Fazekas1999}
\begin{eqnarray}\label{subeq:HubbardHamiltonian}
\lefteqn{\hat{H}_{\rm MFAH}  =  -t \sum_{\langle {\bf i,j} \rangle} \sum_{\sigma = {\uparrow,\downarrow}}{\left(\hat{c}_{{\bf i}\sigma}^{\dag}\hat{c}_{{\bf j}\sigma}+\hat{c}_{{\bf j}\sigma}^{\dag}\hat{c}_{{\bf i}\sigma}\right)}} \nonumber \\
&&{} + U \sum_{\bf i} \left\{\hat{c}_{{\bf i}\uparrow}^{\dag} \hat{c}_{{\bf i}\uparrow} \left(
n_{{\bf i}\downarrow} - \frac{n_{\bf i}}{2}
\right)+\hat{c}_{{\bf i}\downarrow}^{\dag}\hat{c}_{{\bf i}\downarrow}\left(
n_{{\bf i}\uparrow}- \frac{n_{\bf i}}{2} \right)\right\} \nonumber \\
&&{} + \sum_{\bf i} \sum_{\sigma = \uparrow,\downarrow} v_{\bf i} \hat{c}_{{\bf i}\sigma}^{\dag} \hat{c}_{{\bf i}\sigma},
\end{eqnarray}
can be estimated by fitting~\cite{Pisani2007} the spin-unrestricted DFT band structure near the Fermi energy $E_F=0$ with that obtained from the MFAH Hamiltonian~(\ref{subeq:HubbardHamiltonian}) defined on \mbox{$N_z$-ZGNR} honeycomb lattice.  The values for $t$ and $U$  estimated in this fashion slightly depend on the choice of the exchange-correlation functional employed within DFT approximation schemes.~\cite{Pisani2007}  In Eq.~(\ref{subeq:HubbardHamiltonian}), operator $\hat{c}_{\bf i}^\dag$ ($\hat{c}_{\bf i}$) creates (annihilates) an electron in the $\pi$-orbital located at site ${\bf i}=(i_x,i_y)$ of the honeycomb lattice. The third term, which is zero for charge neutral systems, accounts for the shift of the on-site energies due to Coulomb interaction with the  applied electric fields or uncompensated charges in the system---the coefficients $v_{\bf i}$ are to be computed self-consistently in a standard DFT-like fashion, as elaborated in  Sec.~\ref{sec:NewtonRaphson}.

As customary,~\cite{Son2006a} the width of \mbox{$N_z$-ZGNR} lattice is measured using the number $N_z$ of zigzag longitudinal chains. The number of atoms $N_a^z$ comprising a single longitudinal zigzag measures its length. In the units of graphene lattice constant $a=2.46$ \AA, the average width of ZGNR is $W=a\sqrt{3}(N_z-1)/2$ and its length is $L=a(N_a^z-1)/2$.

The spin-resolved ($\sigma=\uparrow,\downarrow$ along the $z$-axis orthogonal to ZGNR plane) electron density on carbon atom at site ${\bf i}$ is given by the statistical expectation value
\begin{equation}
n_{{\bf i}\sigma} = \langle \hat{c}_{{\bf i}\sigma}^{\dag} \hat{c}_{{\bf i}\sigma} \rangle,
\end{equation}
so that particle density at the same site is the sum
\begin{equation}
n_{\bf i}=n_{{\bf i} \uparrow}+n_{{\bf i} \downarrow}.
\end{equation}
These quantities have to be computed via the self-consistent loop,~\cite{Ohno2000a} either from the eigenstates of equilibrium systems~\cite{Fernandez-Rossier2007} or from NEGFs (Sec.~\ref{sec:NewtonRaphson}) when finite-length ZGNR is attached to electrodes~\cite{Dragomirova2008a} and brought into nonequilibrium state by the applied bias voltage. Once the self-consistency criterion is satisfied,  the spatial distribution of magnetization density within ZGNR is obtained from
\begin{equation}
m_{\bf i} = g \mu_B S_{\bf i}^z = \mu_B (n_{{\bf i}\uparrow} - n_{{\bf i}\downarrow}),
\end{equation}
where $S_{\bf i}^z$ is the spin density and $\mu_B$ is the Bohr magneton.

We chose to combine local orbital basis Hamiltonian~(\ref{subeq:HubbardHamiltonian}) with NEGF because it allows us to substantially accelerate self-consistent calculations in the nonequilibrium current-carrying state (as discussed in Sec.~\ref{sec:NewtonRaphson}). Although Eq.~(\ref{subeq:HubbardHamiltonian}) is typically obtained through mean-field decoupling scheme~\cite{Fazekas1999} by starting from the full many-body Hubbard model for lattice fermions, it can also be justified with the framework of LSDA. Furthermore,
the latter provides simple and clear explanation of the expression for the total energy, which will be required for thermodynamic analysis of Sec.~\ref{sec:Thermodynamics}.

By using the spin-restricted self-consistent environment-dependent tight-binding model (SC-EDTB), which is specifically tailored to simulate eigenvalue spectra, electron densities and Coulomb potential distributions for  carbon-hydrogen systems,~\cite{Areshkin2004,Areshkin2005} we can establish the relationship between Hamiltonian~(\ref{subeq:HubbardHamiltonian}) and its LSDA counterpart~\cite{Fiolhais2003}
\begin{eqnarray} \label{subeq:LSDAHamiltonian}
\hat{H}_{\rm LSDA}^\sigma &  = & \frac{\hat{\bf p}^2}{2 m} + e^2 \int d^3r^\prime \, \frac{n({\bf r}')}{|{\bf r}-{\bf r}'|} + V_{\rm pp}({\bf r}) + V_{\rm ext}({\bf r}) \nonumber \\
&&{}  +  V_{\rm xc}([n];{\bf r}) + \Delta V_{\rm xc}^\sigma, \\
\Delta V_{\rm xc}^\sigma & = & V_{\rm xc}^{\sigma}\left([n^{\uparrow},n^{\downarrow}];{\bf r}\right) - V_{\rm xc}([n];{\bf r}).
\end{eqnarray}
The first five terms in this one-electron Hamiltonian are: kinetic energy operator, classical Hartree potential, pseudopotential associated with core electrons, external potential, and spin-restricted part of exchange-correlation potential, respectively. They do not depend on the spin polarization and can be accounted by the SC-EDTB model. The EDTB aspect~\cite{Tang1996} of the model assumes that hopping matrix elements of the tight-binding Hamiltonian depend not only on the distance between the two atoms on which the basis functions are centered, but also on the arrangement of neighboring atoms (i.e., it is analogous to a DFT scheme that accounts for three- and four-centre integrals, and with atomic orbitals adjusted to atomic environment). The SC-EDTB model adds parameters to this non-self-consistent EDTB part in order to describe hydrocarbon bonds while taking into account the self-consistent~\cite{Ohno2000a} charge transfer.~\cite{Areshkin2004,Areshkin2005} The last term $\Delta V_{\rm xc}^\sigma$ in Eq.~(\ref{subeq:LSDAHamiltonian}) is different for $\uparrow$ and $\downarrow$  spins, where the spin-dependent $V_{\rm xc}^{\sigma}\left([n^{\uparrow},n^{\downarrow}];{\bf r}\right)$ exchange-correlation potential in LSDA is~\cite{Fiolhais2003}
\begin{equation}
\label{eq:VxcThroughnExc}
\left. V_{\rm xc}^{\sigma}([n^{\uparrow},n^{\downarrow}];{\bf r})=\frac{\partial}{\partial{n_{\sigma}}} \left[ (n_\uparrow+n_\downarrow) e_{\rm xc}(n_\uparrow,n_\downarrow)\right]\right|_{n_\sigma = n_\sigma(\bf r)}.
\end{equation}
The exchange-correlation energy per particle $e_{\rm xc}(n_\uparrow,n_\downarrow)$ is extracted~\cite{Fiolhais2003} from an electron gas with uniform densities $n_\uparrow$, $n_\downarrow$.

\begin{figure}
\includegraphics[scale=0.6,angle=0]{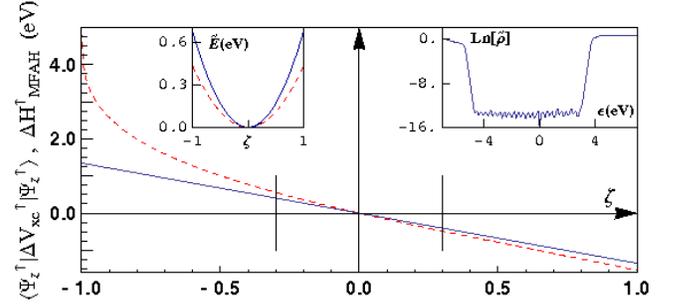}
\caption{(Color online) The spin-dependent contribution to the on-site matrix elements of LSDA and MFAH Hamiltonians as the function of relative spin polarization $\zeta$. The dashed red line plots the expectation value $\langle\Psi_{z}^\uparrow |\Delta V_{\rm xc}^{\uparrow}| \Psi_{z}^\uparrow \rangle$ for Perdew-Zunger parametrization~\cite{Fiolhais2003} and SC-EDTB orbital parameters, where electron density contribution from other $sp^{2}$ orbitals to $\Delta V_{\rm xc}^{\uparrow}$ is neglected. The solid blue line plots spin-dependent contribution $H_{\rm MFAH}^{\uparrow}[\zeta]$ to the on-site matrix element of MFAH Hamiltonian with $U = 2.7$ eV. The two vertical lines indicate variation range for the spin polarization parameter $\zeta$ within ZGNR.  The right inset plots the logarithm of SC-EDTB $sp^{2}$ DOS as a function of energy ($E_F=0$ is the Fermi level).  The left inset plots spin-dependent contribution $\tilde{E}$ (dashed red)  to the total energy Eq.~(\ref{eq:ETotalLDSA}) in LSDA  and  $-U\sum_{\bf i}{n_{{\bf i}\uparrow}n_{{\bf i}\downarrow}}$ (solid blue) contribution to the total energy Eq.~(\ref{eq:ETotalHubbard}) in MFAH model as a function of $\zeta$.}\label{fig:LSDAvsHubbard}
\end{figure}

Let us compute the on-site $\pi$-orbital matrix element for $\Delta V_{\rm xc}^\sigma$.  We borrow the orbital parameters from SC-EDTB, particularly
\begin{equation}\label{eq:POrbital}
\Psi_z(r,\theta)=\alpha  e^{-a_{p}r^{2}} r \cos \theta,
\end{equation}
which is one of the four localized orbitals per carbon atom comprising the basis set. Here $a_{p}=1.6085$ {\AA}$^{-1/2}$ and $\alpha$ is the
normalization factor. By defining local relative spin polarization
\begin{equation}\label{eq:polarization}
\zeta({\bf r})=\frac{n^{\uparrow}({\bf r})-n^{\downarrow}({\bf r})}{n^{\uparrow}({\bf r})+n^{\downarrow}({\bf r})},
\end{equation}
and by assuming that within the orbital range $\zeta$ does not depend on ${\bf r}$, the spin-resolved electron densities for $\pi$-orbital are
\begin{equation}\label{eq:SpinDensityThroughSpinPolarizationParameter}
n^{p_{z}}_{\uparrow}({\bf r})=
\Psi_{z}^{2}\frac{1+\zeta}{2}, \ n^{p_{z}}_{\downarrow}({\bf r})=
\Psi_{z}^{2}\frac{1-\zeta}{2}.
\end{equation}
Equations~(\ref{eq:VxcThroughnExc}) and ~(\ref{eq:POrbital}) determine the matrix element  $\langle\Psi_{z}^\sigma|\Delta V_{\rm xc}^{\sigma}|\Psi_{z}^\sigma\rangle$ as a function
of $\zeta$ ($\Psi_{z}^\sigma=\Psi_{z}\chi_\sigma$ where $\chi_\sigma$ is the spinor part of the wave function).

Similarly, we can obtain the spin-dependent contribution to on-site matrix elements of MFAH Hamiltonian~(\ref{subeq:HubbardHamiltonian})
\begin{equation}\label{eq:DeltaHHubbard}
\Delta H_{\rm MFAH}^{\sigma}[\zeta]= \mp \frac{U}{2} n_{\bf i} \zeta \approx \mp \frac{U}{2} \zeta,
\end{equation}
which is a linear function of $\zeta$. As we demonstrate below, even in the strong electric field the relative change of the total $\pi$-orbital electron population does not exceed 1$\%$.  This means that $n_{\bf i}$ in Eq.~(\ref{eq:DeltaHHubbard}) can be assumed equal to unity.

Figure~\ref{fig:LSDAvsHubbard} plots the spin-dependent part of the on-site matrix elements of LSDA and MFAH Hamiltonians.  Since in
ZGNR systems the polarization $\zeta$ varies within the interval $[-0.3, +0.3]$, we find a good fit between matrix elements of
MFAH and LSDA Hamiltonians in this range, thereby justifying the usage of Eq.~(\ref{subeq:HubbardHamiltonian}) instead of
more complicated Eq.~(\ref{subeq:LSDAHamiltonian}). The fit also sets the parameters $t = U = 2.7$ eV for
the MFAH model. The right inset in \mbox{Fig.~\ref{fig:LSDAvsHubbard}} plots the \mbox{SC-EDTB} contribution of
\mbox{$sp^{2}$} carbon orbitals to the density of states (DOS) for a typical single-layer graphene system composed of GNRs of different types and widths.  As follows from the inset, the orbitals other than $p_{z}$ do not have any contribution to the DOS within $[-4 \ {\rm eV}, +3 \ {\rm eV}]$
interval around the Fermi energy. This suggests that usage of single \mbox{$\pi$-orbital} per honeycomb lattice site in Hamiltonian~(\ref{subeq:HubbardHamiltonian})
should be sufficient in the cases when the applied bias voltage does not exceed $\pm 2$~V.

From this analysis, as well as from the mappings~\cite{Fernandez-Rossier2007,Pisani2007} of DFT calculations to simpler MFAH model or the fact that DFT results obey
the Lieb theorem~\cite{Lieb1989} for the exact ground state of the Hubbard model on charge neutral bipartite lattices, we can conclude that second neighbor hopping and
intersite Coulomb repulsion (present in the DFT calculations) do not modify the relation between lattice imbalance and total spin of the ground state
warranted for the Hubbard model for which these couplings are absent. Thus, given that the replacement of LSDA by MFAH Hamiltonian is reasonably
justified, we proceed to derive expression for the total energy as a function of $\zeta$ based on the solution of Eq.~(\ref{subeq:HubbardHamiltonian}). The total
energy within the LSDA framework is
\begin{subequations}
\begin{eqnarray}
E^{\rm Total}_{\rm LSDA} [n,\zeta]  & = &   \sum_{i} {f(\varepsilon_i-\mu) \varepsilon_i} + \sum_{i<j}^{N_{\rm atoms}}{ \frac{e^2 Z_{i} Z_{j}}{\mid {\bf R}_{i}-{\bf R}_{j}\mid}} \nonumber \\
&&{} + \frac{e^2}{2}\int\!\!\int d^3r d^3r' \, \frac{n({\bf r})n({\bf r}')}{\mid {\bf r}-{\bf r}'\mid } \nonumber \\
&&{} + \tilde{E} [n,\zeta],\label{subeq:ETotal} \\
\tilde{E} [n,\zeta] &  = & \sum_{\sigma=\uparrow,\downarrow} \int d^3 r \, n_{\sigma}({\bf r}) V_{\rm xc}^{\sigma}([n,\zeta];{\bf r}) \nonumber \\
&&{} + \int d^3r \, n({\bf r}) e_{\rm xc}(n_\uparrow({\bf r}),n_\downarrow({\bf r})).
\label{subeq:ETildaThroughnUpnDnexc}
\end{eqnarray}\label{eq:ETotalLDSA}
\end{subequations}
Here $\varepsilon_i$ are the eigenvalues of Eq.(\ref{subeq:LSDAHamiltonian}), $f(\varepsilon)$ is the Fermi function, $\mu$ is the chemical potential [chosen to satisfy $\int d^3r\, n({\bf r})=N$ where $N$ is the total number of electrons], $Z_{i}$ are atomic core charges, and ${\bf R}_{i}$ are nuclear coordinates.  The self-consistent computation of spin and particle densities shows that transition between the spin-polarized and the spin-restricted solution does not result in any noticeable change in the total electron density.  We also assume the same atomic coordinates for the entire range of interest for $\zeta$.  Therefore, the second and the third term in Eq.(\ref{subeq:ETotal}) do not depend on spin polarization.  If we assume that $\zeta$ is uniform, i.e., independent of ${\bf r}$ within single orbital range, then by substituting the electron densities Eq.~(\ref{eq:SpinDensityThroughSpinPolarizationParameter}) into $\tilde{E}[n,\zeta]$ we obtain nearly quadratic dependence of $\tilde{E}$ on $\zeta$, as shown in the left inset (dashed line) of Fig.~\ref{fig:LSDAvsHubbard}.

The expression for the total energy within the MFAH model is given by~\cite{Gunlycke2007}
\begin{equation}\label{eq:ETotalHubbard}
E^{\rm Total}_{\rm MFAH}[n,\zeta]= \sum_{i} f(\varepsilon_i-\mu) \varepsilon_i - U \sum_{\bf i} n_{{\bf i}\uparrow}n_{{\bf i}\downarrow}.
\end{equation}
The second term in Eq.~(\ref{eq:ETotalHubbard})  is plotted (solid line) as a function of $\zeta$ in the left inset of Fig.~\ref{fig:LSDAvsHubbard} and represents approximation for $\tilde{E}$ in Eq.~(\ref{eq:ETotalLDSA}) plotted in the same inset (dashed line).  As demonstrated by Fig.~\ref{fig:LSDAvsHubbard}, the LSDA matrix element averaged over the range \mbox{$\zeta \in [-0.3,0.3]$} is positive due to slightly superlinear dependence on $\zeta$.  At the same time the average of the Hubbard matrix element over the same interval is exactly zero due to its linear dependence on $\zeta$.  That is, on average MFAH model underestimates the on-site Hamiltonian matrix elements, but it overestimates $\tilde{E}$.  The partial error compensation in Eq.~(\ref{eq:ETotalHubbard}) makes $U=2.7$ eV a reasonable choice for approximation of both the LSDA single particle energies and the total energy by a simpler MFAH model.

The energy expression~(\ref{eq:ETotalHubbard}) also provides a transparent explanation for the origin of magnetic ordering in ZGNRs. The value of $\zeta = -1$ in Fig.~\ref{fig:LSDAvsHubbard} corresponds to spin-$\uparrow$ electron surrounded by spin-$\downarrow$ electron density, while $\zeta = 1$ is associated with spin-$\uparrow$ electron surrounded by spin-$\uparrow$ electron density.  Therefore, the Hamiltonian matrix elements favor the spin polarization.  At the same time, the second term in Eq.~(\ref{eq:ETotalHubbard}) favors the non-spin-polarized solution (see left inset in Fig.~\ref{fig:LSDAvsHubbard}), so that the competition between this term proportional to $\zeta^{2}$ and the band energy proportional to $\zeta$ determines the appearance of non-zero spin polarization.

\section{NEGF with Accelerated Convergence Self-Consistent Algorithm} \label{sec:NewtonRaphson}

We employ the NEGF formalism~\cite{Haug2007}  for the computation of nonequilibrium  spin-resolved electron densities by starting from the MFAH Hamiltonian Eq.~(\ref{subeq:HubbardHamiltonian}). Assuming a two-terminal system, composed of finite-size ZGNR attached via semi-infinite ideal leads to the left ($L$) and right ($R$)
macroscopic reservoirs where electrons thermalize to be characterized by the electrochemical potentials $\mu_{L} >\mu_{R}$, the nonequilibrium electron density
\begin{equation}\label{eq:chargevector}
{\bf n}={\tt diag}\, [{\bf D}],
\end{equation}
in the phase-coherent approximation (i.e., in the absence of dephasing and inelastic processes~\cite{Golizadeh-Mojarad2007a}) is obtained from the following density matrix
\begin{eqnarray}
\label{eq:NEGFDensity}
\lefteqn{{\bf D}  =  -\frac{1}{\pi}\int\limits_{-\infty}^{+\infty}dE \, {\rm Im}\, [{\bf G}(E)] f(E-\mu_{R})} \nonumber\\
&&{} - \frac{1}{\pi}\int\limits_{-\infty}^{+\infty}dE\, {\rm Re}\, \left\{ {\bf G}(E) {\rm Im}\, [{\bm \Sigma}_{L}(E)] {\bf G}^{\dagger}(E) \right\} \nonumber\\
&&{} \times \left[f\left(E-\mu_{L}\right)-f\left(E-\mu_{R}\right)\right],
\end{eqnarray}
Here ${\bf G}$ is the retarded Green function matrix and ${\tt diag}\, [\ldots]$ returns vector composed of the diagonal elements of its argument. Because Eq.~(\ref{subeq:HubbardHamiltonian}) assumes zero overlap between the orbitals, only diagonal elements of ${\bf D}$ contribute to electron density in Eq.~(\ref{eq:chargevector}). The retarded self-energy matrix ${\bm \Sigma}_{L}$ is introduced by the ``interaction'' with the left lead---it determines escape rates of electrons into the left reservoir.~\cite{Haug2007}
The density matrix in Eq.~(\ref{eq:NEGFDensity}) is split into equilibrium (first term) and nonequilibrium (second term) contributions,~\cite{Taylor2001,Brandbyge2002,Ke2004} taking into account that left-lead states are filled up to $\mu_L$ and right-lead states are filled up to energy $\mu_R$. Integration over energy is performed using the poles summation algorithm.~\cite{Areshkin2009}

For a small difference between $\mu_{L}$ and $\mu_{R}$, the self-consistency can be achieved by applying the Broyden convergence acceleration method,~\cite{Ohno2000a,Ihnatsenka2007,Marks2008} which has two major advantages.  First, the Broyden method is compatible with the recursive algorithm for construction of the Green functions and self-energies, where recursion is extended to allow for the computation of local quantities inside the sample rather than usual transmission function and conductance determined by it.~\cite{Cresti2003,Metalidis2005,Kazymyrenko2008}  The simplest version of such algorithms starts  by partitioning the quasi-linear system into slices (described by a much smaller Hamiltonian matrix) in a such way that only the coupling between the nearest neighbor slices is present.  Then, the recursive algorithm is applied to propagate the self-energies from the contacts and to build the Green functions for each slice.  The nonequilibrium electron density for each slice is derived locally from the Green functions and the self-energies for the given slice.  The computation time scales linearly with the number of slices and cubically with the size of the matrices (Hamiltonian, Green functions, and self-energies) associated with a single slice. Second, the Broyden method adds $O(N)$ extra operations and hence does not slow down the single iteration.  However, the reduction of the iteration number achieved by the Broyden method is appreciable.  For equilibrium problems considered in Sec.~\ref{sec:Thermodynamics} the number of iterations required to achieve $10^{-10}$ maximum difference between the input and output electron densities using the Broyden method is about 30, while the scalar charge mixing requires several hundreds of iterations.~\cite{Ihnatsenka2007}

The Broyden method works well when the correlation between the electron density and the potential is local, i.e., when the local potential distortion results in a local self-consistent density change.  Conversely, in the case of non-local correlations the Broyden method performance rapidly deteriorates.  The nonequilibrium electron density in the coherent ballistic approximation constitutes the perfect example when the Broyden method fails.  The reason for this is that electron-potential correlations becomes completely non-local: the change of the potential at one contact can shut off the electron flux through the entire system and cause the system-wide electron density redistribution. The ``brute-force'' alternative to the Broyden algorithm is the Newton-Raphson method, which slows down each iteration by an order of magnitude, but reduces the number of iterations to less than a dozen and guarantees the convergence towards the self-consistent solution.

Because the convergence under nonequilibrium conditions constitutes the major computational problem, we present the details of the NEGF-adapted Newton-Raphson method employed in our study.  The first order Taylor expansion for the retarded Green function (${\bm \Sigma}={\bm \Sigma}_L + {\bm \Sigma}_R$)
\begin{equation}\label{eq:retardedgreen}
{\bf G}(E) = [E + {\bf H}+{\bm \Sigma}]^{-1},
\end{equation}
with respect to the Hamiltonian variation $\delta {\bf H}$ is
\begin{eqnarray}
\label{eq:TaylorExpansion}
\delta {\bf G}(E) & = & [ E  - ({\bf H} + \delta {\bf H})-{\bm \Sigma} ]^{-1}  - [E - {\bf H} - {\bm \Sigma}]^{-1} \nonumber \\
& = & {\bf G} \cdot {\bf \delta H} \cdot {\bf G}.
\end{eqnarray}
At this point the minimal-basis-set of the MFAH model comes into play---according to Eq.(\ref{subeq:HubbardHamiltonian}) only diagonal matrix elements are affected by the electron density distribution.  In the following, $\delta {\bf H}$ denotes the change of the Hamiltonian due to a small variation of the electron density, which means that $\delta {\bf H}$ is a diagonal matrix.  Vector  $\delta {\bf h}$ denotes the diagonal elements of $\delta {\bf H}$.  Using the symmetry of the retarded Green function matrix associated with the real Hamiltonian, the variation of the electron density $\delta {\bf n}$ with respect to $\delta {\bf h}$ can be written as (all quantities depend on energy $E$ which is omitted for brevity):
\begin{subequations}
\begin{eqnarray}
\delta {\bf n} & = & {\bf A} \cdot \delta {\bf h},\label{subeq:ElectronDensityVariationGeneral}
\\
\nonumber\\
{\bf A} & = & -\frac{1}{\pi} \int\limits_{-\infty}^{+\infty}  dE\, {\rm Im} \left[{\bf G} \otimes {\bf G}\right] f(E-\mu_{R}) \nonumber \\
&&{} - \frac{2}{\pi} \int_{-\infty}^{+\infty} dE\, {\rm Re} \left\{{\bf G}^{\dagger} \otimes \left({\bf G} \cdot {\rm Im} [{\bm \Sigma}_{L}] \cdot {\bf G}^{\dagger}\right) \right\} \nonumber \\
&&{} \times [f(E-\mu_{L})-f(E-\mu_{R})].
\label{subeq:ElectronDensityVariationMtrxADefinition}
\end{eqnarray}\label{eq:ElectronDensityVariation}
\end{subequations}
Here the symbol $\otimes$ between matrices denotes element-wise product of two matrices, so that the element of, e.g., ${\bf G} \otimes {\bf G}$ is $(G_{pq})^2$.  The computational complexity of the integrand in Eq.~(\ref{eq:ElectronDensityVariation}) is $O\left(N^{3}_{\bf G}\right)$ per energy point, where $N_{\bf G}$ is the size of matrix ${\bf G}$.

In the spin-unrestricted case the electron density vector is composed of ${\bf n}_{\uparrow}$ and ${\bf n}_{\downarrow}$ sub-vectors. Therefore, we can rewrite Eq.~(\ref{eq:ElectronDensityVariation}) as a matrix equation:
\begin{eqnarray}
\label{eq:SpinPolarizedDensityVariation2}
\left(
\begin{array}{c}
\delta {\bf n}_{\uparrow}\\
\delta {\bf n}_{\downarrow}
\end{array}\right)=\left(
\begin{array}{cc}
 {\bf A}^{\uparrow} & 0 \\
0 & {\bf A}^{\downarrow}
\end{array}\right)\cdot\left(
\begin{array}{c}
\delta {\bf h}_{\uparrow}\\
\delta {\bf h}_{\downarrow}
\end{array}\right).
\end{eqnarray}
Matrices ${\bf A}^\sigma$ are computed using Eq.~(\ref{subeq:ElectronDensityVariationMtrxADefinition}) with Green function matrices ${\bf G}^{\sigma\sigma}$ plugged in. The matrices ${\bf G}^{\sigma\sigma}$ are obtained by inverting via Eq.~(\ref{eq:retardedgreen}) the corresponding diagonal block ${\bf H}^\sigma$ (for spin-$\sigma$ electrons) of the matrix representation of Hamiltonian~(\ref{subeq:HubbardHamiltonian}).  We assume that there is no spin polarization in the leads, so that the self-energies ${\bm \Sigma}_{L}$ and  ${\bm \Sigma}_{R}$ are the same for both spin polarizations.

In the framework of MFAH model, the on-site potential variation vector is the linear function of the density variation vector
\begin{eqnarray}
\label{eq:PotentialvsDensity1}
\left(
\begin{array}{cc}
\delta {\bf h}_{\uparrow}\\
\delta {\bf h}_{\downarrow}
\end{array}\right)=\left(
\begin{array}{cc}
{\bf Q} & {\bf Q} + U{\bf I}\\
{\bf Q} + U{\bf I} & {\bf Q}
\end{array}\right)\cdot\left(
\begin{array}{cc}
\delta {\bf n}_{\uparrow}\\
\delta {\bf n}_{\downarrow}
\end{array}\right).
\end{eqnarray}
Here ${\bf Q}$ is the Coulomb interaction matrix computed for $\pi$-orbital wave functions with SC-EDTB parameters in Eq.~(\ref{eq:POrbital}) using standard DFT approach.  The dot-product
of the $i^{th}$ row of matrix ${\bf Q}$ and uncompensated $\pi$-orbital electron density (${\bf n}-1$) plus the potential shift due to the external electric field equals the coefficient $v_{\bf i}$ in Eq.~(\ref{subeq:HubbardHamiltonian}).  The identity matrix ${\bf I}$ has the same dimensions as ${\bf Q}$ and ${\bf H}^\sigma$. By defining matrix ${\bf B}$ as
\begin{eqnarray}
\label{eq:PotentialvsDensity2}
{\bf B}=\left(
\begin{array}{cc}
{\bf A}^{\uparrow} & 0 \\
0 & {\bf A}^{\downarrow}
\end{array}\right)\cdot
\left(
\begin{array}{cc}
{\bf Q} & {\bf Q}+U{\bf I} \\
{\bf Q} + U{\bf I} & {\bf Q}
\end{array}\right),
\end{eqnarray}
we can relate, in the first order approximation, the response of the output spin-resolved electron density with respect to a small variation of the input density:
\begin{eqnarray}
\label{eq:SpinPolarizedDensityVariation1}
\delta {\bf n}_{\rm out}\equiv\left(
\begin{array}{c}
\delta {\bf n}_{\uparrow}\\
\delta {\bf n}_{\downarrow}
\end{array}\right)_{\rm out} = {\bf B} \cdot
\left(
\begin{array}{c}
\delta {\bf n}_{\uparrow}\\
\delta {\bf n}_{\downarrow}
\end{array}\right)_{\rm in}
\equiv {\bf B} \cdot \delta {\bf n}_{\rm in}.
\end{eqnarray}
If ${\bf n}_{\rm in}$ is the input density of the self-consistent loop, and ${\bf n}_{\rm out}$ is the corresponding output density, the self-consistent solution can be written as
\begin{eqnarray}
\label{eq:SCConditionGeneral}
{\bf n}_{\rm out}+\delta {\bf n}_{\rm out}={\bf n}_{\rm out} + {\bf B} \cdot \delta {\bf n}_{\rm in} = {\bf n}_{\rm in} + \delta {\bf n}_{\rm in}.
\end{eqnarray}
Equation~(\ref{eq:SCConditionGeneral}) allows us to compute $\delta {\bf n}_{\rm in}$ for the next self-consistent iteration from ${\bf n}_{\rm in}$ and ${\bf n}_{\rm out}$ in the current iteration by solving the system of linear equations
\begin{eqnarray}
\label{eq:SCConditionLinearSystem}
({\bf I}_{\bf B}-{\bf B}) \cdot \delta {\bf n}_{\rm in} = {\bf n}_{\rm out} - {\bf n}_{\rm in}.
\end{eqnarray}
Here ${\bf I}_{\bf B}$ is the identity matrix of the same dimension as matrix ${\bf B}$.

The main computational disadvantage of the Newton-Raphson method is that Eq.~(\ref{subeq:ElectronDensityVariationMtrxADefinition}) uses the full retarded Green function matrix ${\bf G}$, rather than its diagonal part as does the Broyden method.  This prohibits the usage of the recursive Green function algorithm, and requires to apply the Newton-Raphson scheme to matrices containing the information about the entire system rather than to much smaller matrices containing the information about its slices.  Given that the second term in Eq.(\ref{subeq:ElectronDensityVariationMtrxADefinition}) must be evaluated for about one thousand different energy poles, we are limited to systems composed of relatively small number of carbon atoms---the largest out-of-equilibrium ZGNR-based two-terminal device treated in Sec.~\ref{sec:NonEquilibriumCase} contains about one thousand atoms.

\section{Equilibrium Thermodynamics of ZGNR} \label{sec:Thermodynamics}

\subsection{Finite-length ideal ZGNR} \label{sec:EquilibriumInfiniteZGNR}

\begin{figure}
\includegraphics[scale=0.6,angle=0]{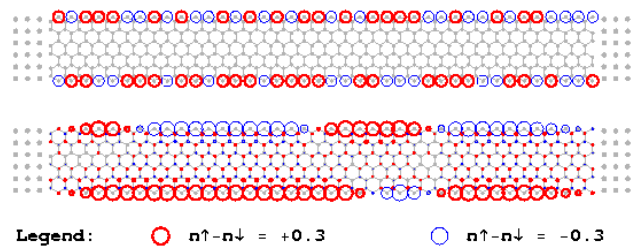}
\caption{(Color online) Bottom panel: Self-consistently computed equilibrium spin density within finite-length \mbox{6-ZGNR} attached to two semi-infinite square lattice leads  at \mbox{$T=293$~K}. Upper panel: The initial random spin density used to obtain the solution in the bottom panel. Thick red and thin blue circles denote spin-$\uparrow$ and spin-$\downarrow$ densities, respectively, with the circle radius being proportional to spin density on the corresponding carbon atom.}\label{fig:RandomPolarization}
\end{figure}

With few exceptions,~\cite{Kim2008,Dragomirova2008a} theoretical investigations of magnetic ordering in ZGNRs have concentrated largely on all-graphitic structures (with addition of different types of edge carbon atom passivation~\cite{Pisani2007}). On the other hand, in experiments, the ultimate electronic contacts are metallic, as illustrated by sub-10-nm-wide GNRFETs with Pd source and drain electrodes.\cite{Wang2008a,Li2008} We first analyze equilibrium magnetic properties of ZGNRs of finite length, with no defects and bounded by perfectly formed zigzag edge, which are attached to metallic leads modeled as semi-infinite square lattice wires. The device setup is illustrated  in Fig.~\ref{fig:RandomPolarization}.  We assume that on the square tight-binding lattice of the leads only the nearest-neighbor hopping $t_{sl}=t=2.7$ eV is different from zero, and the coupling of the leads to central ZGNR sample is described by the same hopping parameter $t_c=t$.

At the Fermi energy ($E_F=0$) of undoped graphene, such leads have maximum number of open transverse propagating modes, which can penetrate into ZGNR as evanescent modes.~\cite{Dragomirova2008a,Robinson2007} In fact, at clean armchair left and right interfaces of ZGNR mode mixing occurs, thereby effectively acting as disorder whose effect on lead-ZGNR contact transparency further depends on weather the lead is ``lattice-matched'' or ``lattice-unmatched'' to the honeycomb lattice of ZGNR.~\cite{Robinson2007} The leads shown in Fig.~\ref{fig:RandomPolarization} fall in the category of ``lattice-unmatched'' ones,~\cite{Dragomirova2008a} as discussed in more detail in Sec.~\ref{sec:FiniteZGNRsquare}.

The two-terminal device in Fig.~\ref{fig:RandomPolarization} is macroscopically inhomogeneous, so that even in equilibrium it is more efficient to use NEGF (with semi-infinite leads accounted through self-energies discussed in Sec.~\ref{sec:NewtonRaphson}) to obtain the texture of its spin polarization,~\cite{Dragomirova2008a} rather than trying to match the eigenstates of the leads to the eigenstates of ZGNR.  If one starts with the random spin polarization illustrated in the upper panel of {Fig.~\ref{fig:RandomPolarization}, the self-consistent solution converges to the magnetization density forming a pattern of finite length segments whose spins are oriented in the same direction.  For example, the lower panel of Fig.~\ref{fig:RandomPolarization} displays one possible self-consistent solution originating from the initial spin density in the upper panel. The magnetization texture within ZGNR lowers the total energy but at the same time it decreases the entropy by aligning electron spins.  Thus, at finite temperature one can expect that spin density along zigzag edges would be represented by finite length segments of a given polarization, as exemplified by the lower pane in Fig.~\ref{fig:RandomPolarization}.

\begin{figure}
\includegraphics[scale=0.7,angle=0]{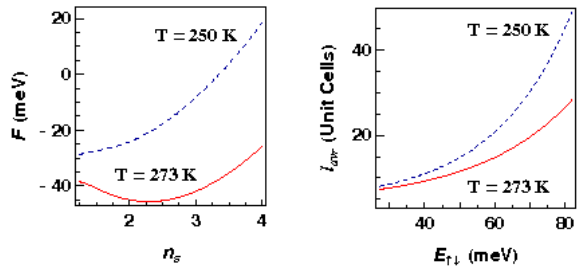}
\caption{(Color online) Left panel: Free energy of the zigzag edge of length $N_z^a=101$ (50 one-dimensional unit cells) as a function of the number of segments with uniform spin-polarization at two different temperatures. The minimum at $n_{s} \approx 2.2$ indicates that at $T = 273$~K the ZGNR edge is partitioned into two segments, while the absence of the minimum at \mbox{$T = 250$~K} means that the number of partitions is less than one so that the whole edge is uniformly spin-polarized.  Right panel: The average length of a uniformly spin-polarized segment as a function of the energy $E_{\uparrow\downarrow}$ associated with the ``boundary'' between two segments of opposite spin polarization.}\label{fig:Thermodynamics}
\end{figure}

To estimate the length of a uniformly spin-polarized segment, we compute the free energy $F$ for  ZGNRs of finite length and find its minimum with respect to the number of such segments. This problem can be formulated as follows. Suppose there is continuous $L$ unit-cells-long zigzag edge, which can have either uniform or fragmented spin polarization.  The
length $l_s$ of the shortest possible fragment for any given value of Hubbard $U$ is known from self-consistent calculations.  Obviously, the edge cannot contain more
than $L/l_s$ fragments.  Every boundary between the two fragments adds additional energy $E_{\uparrow\downarrow}$ to the total energy of the edge, which is also determined from the self-consistent loop. If the number of segments with uniform spin polarization is $n_{s} < L/l_s$, there are ``extra'' $L - n_{s} l_s$ edge carbon atoms that can be distributed between $n_{s}$ segments. Thus, this problem maps onto a question: ``In how many ways can we distribute $L- n_{s} l_s$ indistinguishable spheres among $n_{s}$ distinct boxes?'' Its answer is simply
\begin{equation}\label{eq:NumberOfCombinations}
W\left(n_{s},l_s,L\right)=\frac{(L- n_{s} l_s + n_{s}-1)!}{( L - n_{s} l_s)!(n_{s}-1)!}.
\end{equation}
By applying the Stirling approximation for large factorials,  Eq.~(\ref{eq:NumberOfCombinations}) can be transformed into
\begin{eqnarray}\label{eq:NumberOfCombinationsStirlingApprox}
\lefteqn{W\left(n_{s},l_s,L\right)  = } \nonumber \\
&&{} \frac{1}{\sqrt{2\pi}} \left(L+n_{s}(1-l_s)-1\right)^{n_{s}(1-l_s)+L-\frac{1}{2}} \nonumber \\
&&{} \times \left(n_{s}-1\right)^{\frac{1}{2}-n_{s}}\left(L-n_{s} l_s\right)^{n_{s} l_s - L - \frac{1}{2}},
\end{eqnarray}
so that the entropy related to the spin arrangement along the edges is $S = k_B \ln \left[W\left(n_{s},l_s,L\right)\right]$. The free energy is then given by
\begin{eqnarray}\label{eq:FreeEnergy}
\lefteqn{F \left(n_{s},l_s,T,E_{\uparrow\downarrow}\right) =  } \nonumber \\
&&{} \left(n_{s}-1\right)E_{\uparrow\downarrow}-k_B T \ln\left[W\left(n_{s},l_s,L\right)\right].
\end{eqnarray}
The number of segments $n_{s}$ in ZGNR which is $L$ unit-cells-long at equilibrium is obtained from the condition
\begin{equation}\label{eq:FreeEnergyDerivative}
\frac{\partial F}{\partial n_{s}}=0.
\end{equation}
\begin{figure}
\includegraphics[scale=0.64,angle=0]{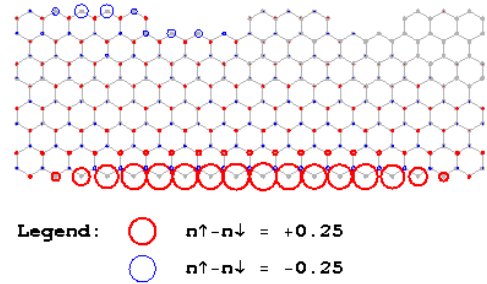}
\caption{(Color online) Equilibrium spin density within finite-length \mbox{8-ZGNR}, whose upper zigzag edge is eroded, at \mbox{$T=293$ K}. The ZGNR is attached to two semi-infinite square lattice leads of the same type as in Fig.~\ref{fig:RandomPolarization}. The dangling bonds of edge atoms are assumed to be passivated with hydrogen atoms (not show explicitly).  The circle radius is proportional to spin polarization at a given atomic site. Thick red circles mark spin-$\uparrow$ density and thin blue circles are for spin-$\downarrow$ density.}\label{fig:ErrodedEdge}
\end{figure}

We use the self-consistent calculations, similar to the one displayed in Fig.~\ref{fig:RandomPolarization} but for longer ZGNRs, to extract the values for $l_s$ and $E_{\uparrow\downarrow}$.  The average number of atoms substantially affected by the transition between the two zigzag edge segments of opposite spin polarization is $\simeq 4$, as illustrated by the lower panel in Fig.~\ref{fig:RandomPolarization}.  The energy of the ``boundary'' between two oppositely spin-polarized edge segments is computed from
\begin{eqnarray}
\label{eq:Theta}
E_{\uparrow\downarrow}=\frac{E_{\rm Total}-E_{\rm Total}^{0}}{N_{b}}.
\end{eqnarray}
The value of $E_{\uparrow\downarrow}$ slightly depends on the ZGNR width and spin ordering type at low temperatures, which can be:~\cite{Pisani2007} (i) {\em antiferromagnetic} (AF), when spin moments on one edge are antialigned to the spin moments on the opposite edge; and (ii) {\em ferromagnetic} (FM), when spin moments on carbon atoms on both edges point in the same direction. Both the AF and FM configurations of magnetic moments have total energy lower than the nonmagnetic state. Moreover, the AF configuration is the ground state in narrow ribbons, while the energy difference between the AF and FM states diminishes with increasing ZGNR width.~\cite{Pisani2007} Here $E_{\rm Total}$ is the total energy of ZGNR with arbitrary fragmented sections of uniform spin polarization along the zigzag edge, $E_{\rm Total}^{0}$ is the total energy of the same ZGNR in the AF ground state, and $N_{b}$ is the number of transitions between $\uparrow$ and $\downarrow$ polarizations on  both edges of ZGNR.

The left panel of \mbox{Fig.~\ref{fig:Thermodynamics}} plots $F$ vs. the number of uniformly spin-polarized segments $n_{s}$ for ZGNR of length $N_z^a=101$ assuming \mbox{$l_s=4$} and \mbox{$E_{\uparrow\downarrow}=80$~meV} ($E_{\uparrow\downarrow}$ can be estimated from the energy gap of ZGNR in magnetic insulating state, see Fig.~\ref{fig:InfiniteStripBandGapCollapse}).  The minimum for the free energy at $T=273$~K indicates that the edges of ZGNR will be most likely partitioned into two uniformly spin-polarized segments but with antialigned spins between the two segments. On the other hand, at \mbox{$T=250$~K} the free energy does not have minimum, meaning that the whole zigzag edge is now uniformly spin-polarized and the corresponding ZGNR is in the AF configuration.  In the limiting case of an infinitely long ZGNR ($L\rightarrow\infty$, $n_{s}\rightarrow\infty$, and $L/n_{s}=l_{\rm avr}$), Eq.~(\ref{eq:FreeEnergyDerivative}) simplifies to
\begin{eqnarray}
\label{eq:AverageContinuousLength}
\exp\left(-\frac{E_{\uparrow\downarrow}}{k T}\right)=\frac{\left(l_{\rm avr}+1\right)^{l_s-1}}{l_{\rm avr}^{l_s}},
\end{eqnarray}
where $l_{\rm avr}$ is the average number of carbon edge atoms in the segment with uniform spin-polarization. The right panel in \mbox{Fig.~\ref{fig:Thermodynamics}} plots $l_{\rm avr}$  as a function of energy $E_{\uparrow\downarrow}$.  For \mbox{$E_{\uparrow\downarrow}=80$~meV}, this length is $l_{avr}\approx 28$ at {$T=273$~K}, while it increases by a factor of two at $T=250$~K.

\subsection{Finite-length ZGNR with disordered edges} \label{sec:EquilibriumInfiniteZGNReroded}

Since edge imperfections are expected to disrupt~\cite{Wimmer2008} the magnetic ordering within ZGNRs, we plot in Fig.~\ref{fig:ErrodedEdge} the self-consistent solution of NEGF-MFAH equations for ZGNR with vacancies along one of its zigzag edges. The removal of a single zigzag chain fragment from ZGNR edge results in almost complete loss of correlations between magnetic moments of edge atoms belonging to different chains. This  means that the length of a segment of edge carbon atoms carrying magnetic moments aligned in the same direction is determined by either topological disorder or thermodynamic disorder, whichever has the smaller characteristic length.

\section{ZGNR in Nonequilibrium Steady-State} \label{sec:NonEquilibriumCase}

\begin{figure}
\includegraphics[scale=0.6,angle=0]{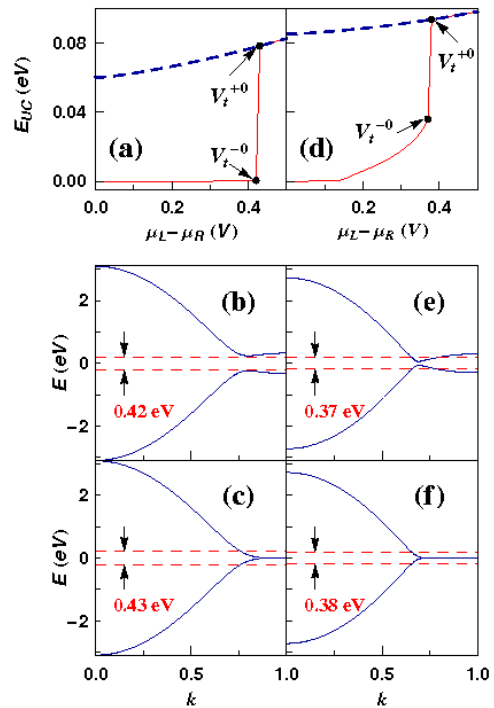}
\caption{(Color online) Total nonequilibrium energy [(a) and (d)] and band structure [(b), (c), (e), and (f)] of infinitely long ideal \mbox{6-ZGNR} [(a)--(c)] and \mbox{32-ZGNR} [(d)--(f)] as the function of applied bias voltage $\mu_L-\mu_R=eV_{ds}$.  Panel (a) plots the total energy per unit cell of \mbox{6-ZGNR} for magnetically ordered AF configuration (solid red) and nonmagnetic state (dashed blue). The threshold voltage at which spin-polarized solution becomes unstable is labeled by $V_{t}$. For $V_{ds}>V_{t}$ the only self-consistent solution available is the non-polarized metallic state. The threshold voltage decreases for wider \mbox{32-ZGNR} in panel (d). Panels (b), (c), (e), and (f) plot the band structure  for the bias voltage slightly below ($V^{-0}_{t}$) and slightly above ($V^{+0}_{t}$) the threshold voltage.  Only two subbands in the vicinity of the Fermi level are shown for clarity, while other subbands experience only minor changes when the transition between spin-polarized and non-polarized states of ZGNR takes place.}\label{fig:InfiniteStripBandGapCollapse}
\end{figure}

\subsection{Infinite ideal ZGNR} \label{sec:InfiniteZGNR}
As it has been previously suggested,~\cite{Gunlycke2007} passing a sufficiently large current along {\em infinitely-long} translationally invariant ZGNR can destroy completely its edge magnetic ordering.  To estimate the source-drain bias voltage
\begin{equation}
eV_{ds}=\mu_{L}-\mu_{R},
\end{equation}
necessary to wash out the spin density, two separate Fermi levels $\mu_{L}$ and $\mu_{R}$ (for the left- and right-moving electrons, respectively) have to be used.  Since this system is infinite and homogeneous, the nonequilibrium spin-resolved electron density on carbon atom at site ${\bf i}$ can be computed simply by using its propagating Bloch modes
\begin{eqnarray}\label{eq:NonEquilibriumDensity}
n_{{\bf i}\sigma} & = & \frac{1}{2} \int dk \sum_{m} |C_{{\bf i}\sigma}^{m}(k)|^{2} \nonumber \\
&&{} \times [f(\varepsilon_{m,k} - \mu_{L}) + f(\varepsilon_{m,k}-\mu_{R})],
\end{eqnarray}
where $C_{{\bf i}\sigma}^{m}(k)$ is the value of the Bloch amplitude obtained by solving Eq.~(\ref{subeq:HubbardHamiltonian}) and the sum over $m$ goes over all bands.  The values of $\mu_{L}$ and $\mu_{R}$ are determined from the charge neutrality condition
\begin{eqnarray}\label{eq:EqForFermiLevel}
\sum_{\bf i} (n_{{\bf i}\uparrow}+n_{{\bf i}\downarrow}) = N_{\rm UC},
\end{eqnarray}
where $N_{\rm UC}$ is the total number of electrons in the unit cell of ZGNR.

Figure~\ref{fig:InfiniteStripBandGapCollapse} depicts the dependence of the total energy and band structure of \mbox{6-ZGNR} and \mbox{32-ZGNR} with respect to the difference between the Fermi levels of the left- and right-moving electrons.  Panels (b)--(f) pertain to the nonequilibrium case, but are similar to equilibrium band structure and can be used to demonstrate the difference between magnetically ordered AF configuration and nonmagnetic state of ZGNRs.  The spin polarization lowers the Hamiltonian eigenenergies, which results in the down-shift of the highest filled band in panels (b) and (e) with respect to non-polarized band structures plotted in panels (c) and (f). Note that both in equilibrium~\cite{Pisani2007} and nonequilibrium studied here, magnetic effects on higher subbands  are negligible, so that Fig.~\ref{fig:InfiniteStripBandGapCollapse} shows only the two subbands around the Fermi energy corresponding to spin-polarized  [Figs.~\ref{fig:InfiniteStripBandGapCollapse}(b) and ~\ref{fig:InfiniteStripBandGapCollapse}(e)] or non-polarized [Figs.~\ref{fig:InfiniteStripBandGapCollapse}(c) and ~\ref{fig:InfiniteStripBandGapCollapse}(f)] edge states.

When the separation between $\mu_{L}$ and $\mu_{R}$ exceeds the certain threshold value, the abrupt change from spin-polarized AF configuration to non-polarized state occurs, as demonstrated by Figs.~\ref{fig:InfiniteStripBandGapCollapse}(a) and ~\ref{fig:InfiniteStripBandGapCollapse}(d).  The step-like transition can be explained as follows.  Suppose the strip is in the AF spin-polarized state and $\mu_{L} > \mu_{R}$.  As the bias voltage is increased, the left-moving electrons with majority spin orientation lying above $\mu_{R}$ are depopulated (i.e., exit through the left contact without being replaced by electrons from the right contact), and the right-traveling  states with minority spin orientation below $\mu_{L}$ are populated.  This reduces the spin density and, hence, decreases the energy  gap in the subband structure plotted in Figs.~\ref{fig:InfiniteStripBandGapCollapse}(b) and ~\ref{fig:InfiniteStripBandGapCollapse}(e). When the device has reached a steady state, the result of these processes can be viewed as an effective repopulation of the electronic states in ZGNR, where electrons are ``excited'' from the valence band into the conduction band, with each such ``excitation'' depopulating a state in the valence band and populating a corresponding state with opposite spin in the conduction band. The reduced spin polarization decreases the band gap, thereby facilitating ``excitations'' that  depopulate more left-traveling states with majority spin orientation lying above $\mu_{R}$ and populate more right-traveling states with minority spin orientation lying below $\mu_{L}$. When separation between $\mu_{L}$ and $\mu_{R}$ exceeds threshold of \mbox{$\approx 0.4$~eV}, this feedback mechanism becomes positive~\cite{Gunlycke2007} and the spin-polarized insulating state collapses to non-polarized metallic solution whose subband structure is shown in Figs.~\ref{fig:InfiniteStripBandGapCollapse}(c) and ~\ref{fig:InfiniteStripBandGapCollapse}(f).

\begin{figure}
\includegraphics[scale=0.6,angle=0]{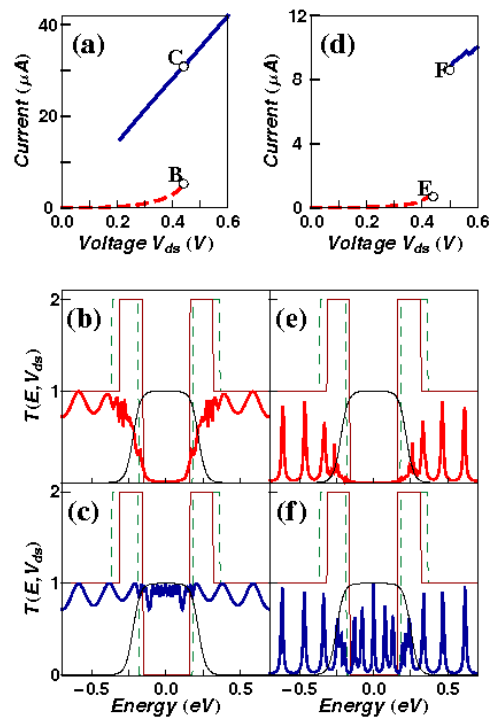}
\caption{(Color online) The \mbox{\em I-V} characteristics [(a) and (d)] and transmission function [(b), (c), (e), and (f)] for \mbox{11-ZGNR} (width $\approx 2.1$ nm; length $\approx 6.6$ nm) two-terminal device depicted in Fig.~\ref{fig:ChargeAndSpinDensity}.  Left column panels are for perfect coupling $t_c=2.7$~eV between ZGNR and multiple-linear-chain leads, while right column panels use $t_c=0.27$~eV which sets the contact resistance to $\approx 60$~k$\Omega$ (as in experiments of Ref.~\onlinecite{Wang2008a}).  In panels (a) and (d), dashed red line denotes \mbox{\em I-V} curves in the AF state of ZGNR, while  solid blue line denotes \mbox{\em I-V} curves for the same  ZGNR after its magnetic ordering is destroyed by the applied bias voltage.  Solid red curve in panels (b) and (e) is transmission function at point {\bf B}  in panel (a) or point {\bf E}  in panel (d), respectively, for nonequilibrium AF spin configuration illustrated in Fig.~\ref{fig:ChargeAndSpinDensity}.  Solid blue curve in panels (c) and (f) is transmission function at point {\bf C} in panel (a) or point {\bf F}  in panel (d), respectively, for  nonmagnetic metallic state. In panels (b), (c), (e), and (f), thin solid black line  illustrates energy window $f(E-eV_{ds}/2)-f(E+eV_{ds}/2)$ over which $T(E,V_{ds})$ is integrated to get the current at corresponding points {\bf B}, {\bf C}, {\bf E}, and {\bf F} in panels (a) and (d), while thin dashed green line and
thin solid red line  are transmission functions of infinitely long ideal \mbox{11-ZGNR} at zero bias $V_{ds}=0$ and threshold bias $V_{ds} = V_{t}^{-0}$ (defined in the same way as in \mbox{Fig. \ref{fig:InfiniteStripBandGapCollapse}}), respectively.}\label{fig:IVMediumSizeStrip}
\end{figure}
\begin{figure}
\includegraphics[scale=0.64,angle=0]{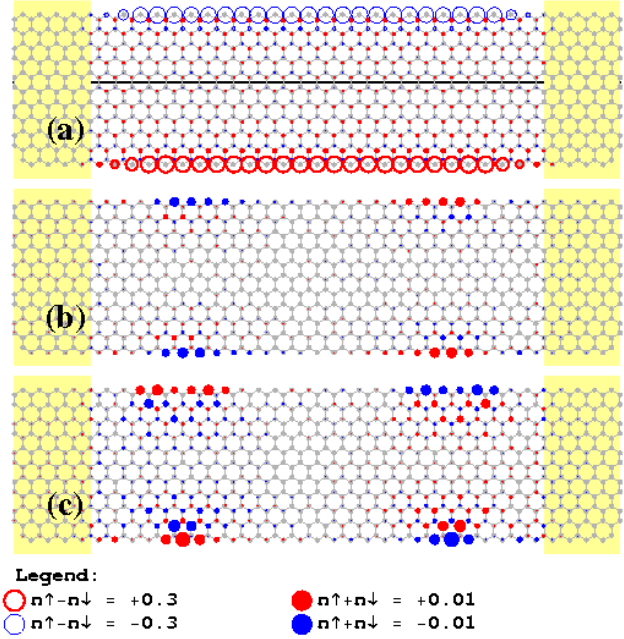}
\caption{(Color online) Spatial profile of spin [panel (a)] and charge [panels (b) and (c)] density within \mbox{11-ZGNR} (width $\approx 2.1$ nm; length $\approx 6.6$ nm) two-terminal device at the threshold voltage $V_{ds}=0.44$~V and for perfect coupling ($t_c=2.7$~eV) to multiple-linear-chain leads attached in yellow-shaded contact regions. In panel (a), the maximum difference between spin-$\uparrow$ and spin-$\downarrow$ electron density is $n_{{\bf i}\uparrow} - n_{{\bf i}\downarrow} \approx 0.24$. The electron density profile in panel (b) corresponds to spatial spin distribution in panel (a) and point {\bf B} in the \mbox{\em I-V} curve in Fig.~\ref{fig:IVMediumSizeStrip}(a). Panel (c) shows spatial profile of electron density  in the nonequilibrium state of ZGNR marked by point {\bf C} in Fig.~\ref{fig:IVMediumSizeStrip}(a), whose spin-polarization is completely washed out.}\label{fig:ChargeAndSpinDensity}
\end{figure}
\begin{figure}
\includegraphics[scale=0.64,angle=0]{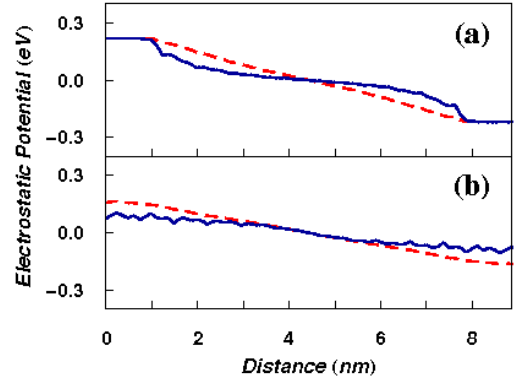}
\caption{(Color online) Electrostatic potential profile along the line drawn in the middle of ZGNR in Fig.~\ref{fig:ChargeAndSpinDensity}(a) for: (a) applied bias voltage \mbox{$V_{ds}=0.44$~V} and perfect coupling \mbox{($t_c=2.7$~eV)} of the contact regions to multiple-linear-chain leads, where dashed red line is for magnetically ordered AF state corresponding to point {\bf B} in Fig.~\ref{fig:IVMediumSizeStrip}(a) and solid blue line corresponds to nonmagnetic metallic state marked by point {\bf C} in Fig.~\ref{fig:IVMediumSizeStrip}(a); (b) coupling $t_c=0.27$~eV and the applied bias voltage $V_{ds}=0.44$~V (dashed red) corresponding to spin-polarized state marked by point {\bf E} in Fig.~\ref{fig:IVMediumSizeStrip}(d), or $V_{ds}=0.5$~V (solid blue) corresponding to nonmagnetic metallic state marked by point {\bf F} in Fig.~\ref{fig:IVMediumSizeStrip}(d).}\label{fig:CoulombPotential}
\end{figure}

\subsection{Finite-length ZGNR attached to multiple-linear-chain leads} \label{subsec:FiniteZGNR}

To simulate metallic contacts on the top of ZGNR, similar to Pd contacts of experimental devices in Ref.~\onlinecite{Wang2008a}, we assume that every carbon atom in the contact regions (shaded with light yellow color in Fig.~\ref{fig:ChargeAndSpinDensity}) is connected to a linear tight-binding chain with hopping parameter between the chain atoms $t_{lc}=2.7$~eV.  Thus, the metallic electrodes of a two-terminal device are simulated with a large number of non-interacting semi-infinite linear chains.  The choice for such model is stipulated by the metallic character of linear chains and convenient way to simulate a top contact connected to a large number of atoms within ZGNR.  Two different hopping parameters between the chains and the carbon atoms are employed: $t_c=2.7$ eV simulates highly transparent contact, while $t_c=0.27$~eV is chosen to correspond to the contact resistance of \mbox{$\simeq 60$~k$\Omega$} measured for GNRFET devices in  Ref.~\onlinecite{Wang2008a}. Also, the width $\sim 2$ nm of all ZGNR we examine below is selected to fall in the range of experimentally fabricated sub-10-nm-wide GNRs,~\cite{Wang2008a} all of which have exhibited semiconducting behavior in the \mbox{\em I-V} characteristics measurements. Note that all-semiconducting nature of ultra narrow GNRs appears to be a key advantage over single-wall carbon nanotubes (which in the similar diameter range are typically a mixture of semiconducting and metallic ones~\cite{Li2008}), as candidates for the envisaged carbon nanoelectronics.~\cite{Avouris2007,Avouris2009}

\begin{figure}
\includegraphics[scale=0.6,angle=0]{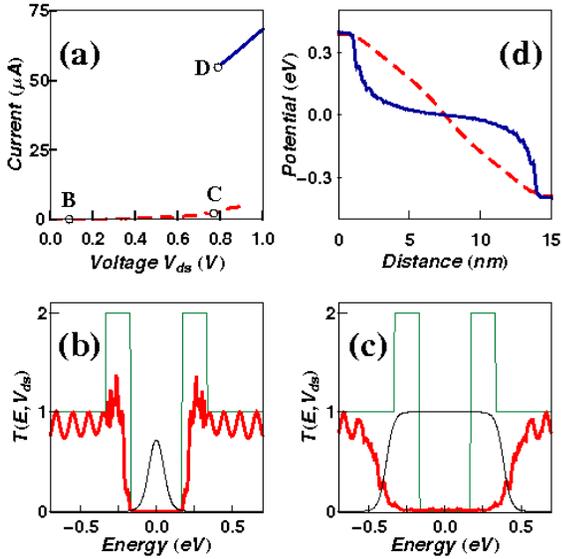}
\caption{(Color online) The (a) \mbox{\em I-V} characteristics, (b)--(c) transmission function, and (d) electrostatic potential profile for \mbox{11-ZGNR} with the distance between the contacts twice as large ($\approx 13.2$ nm) as in the devices in Figs.~\ref{fig:IVMediumSizeStrip} and \ref{fig:ChargeAndSpinDensity}. The coupling between carbon atoms in the contact region and  multiple-linear-chain leads is  perfect \mbox{($t_c=2.7$~eV)}. In panels (a) and (d), dashed red line is for AF state of ZGNR, while solid blue line denotes \mbox{\em I-V} curve (a) or potential profile (d) for ZGNR after its magnetic ordering is destroyed by the applied bias voltage. Point \textbf{B} marks \mbox{$V_{ds}=0.09$~V}, while points \textbf{C} and \textbf{D} correspond to \mbox{$V_{ds}=0.77$~V}. The transmission function (thick red) in panels (b) and (c) is computed at points \textbf{B} and \textbf{C}, respectively, where thin solid black line  illustrates energy window $f(E-eV_{ds}/2)-f(E+eV_{ds}/2)$ over which $T(E,V_{ds})$ is integrated to get the current at corresponding points {\bf B} and {\bf C}. Thin green line is the transmission function of an infinite ideal \mbox{11-ZGNR} at zero applied voltage.  Panel (d) plots the electrostatic potential profile along extended device ZGNR + contact-regions  for spin-polarized AF configuration (dashed red) at  $V_{ds}=0.77$~V  (point \textbf{C}) and non-polarized metallic state (solid blue)  at $V_{ds}=0.77$~V (point \textbf{D}).}\label{fig:IVLongStrip}
\end{figure}

For devices described by an effective single-particle Hamiltonian, such as Eq.~(\ref{subeq:HubbardHamiltonian}), with mean-field treatment of interactions and no dephasing processes the current at finite bias voltage can be computed from the NEGF-based formula~\cite{Haug2007,Taylor2001,Brandbyge2002,Ke2004}
\begin{equation}\label{eq:lb}
I(V_{ds})=\frac{2e}{h} \int\limits_{-\infty}^{+\infty} dE\, T(E,V_{ds}) [f(E-\mu_L)-f(E-\mu_R)],
\end{equation}
which integrates the self-consistent transmission function
\begin{eqnarray}
\lefteqn{T(E,V_{ds})=} \nonumber \\
&&{} {\rm Tr} \left\{ {\bm \Gamma}_R (E+eV_{ds}/2)  {\bf G} {\bm \Gamma}_{L}(E-eV_{ds}/2)  {\bf G}^\dagger \right\},
\end{eqnarray}
for electrons injected at energy $E$ to propagate from the left to the right electrode under the source-drain applied bias voltage $\mu_{L}-\mu_{R}=eV_{ds}$. The energy window for the integral in Eq.~(\ref{eq:lb}) is defined by the difference of Fermi functions  $f(E-\mu_L)-f(E-\mu_R)$ of macroscopic reservoirs into which semi-infinite leads terminate. This ``window'', at room temperature $T=293$~K and for selected bias voltage $V_{ds}$, is shown explicitly (thin black solid line) in Figs.~\ref{fig:IVMediumSizeStrip}, \ref{fig:IVLongStrip}, and \ref{fig:SquareLeads} for three different types of ZGNR devices, where it encloses the portion of $T(E,V_{ds})$ vs. $E$ curve which is integrated to get the current $I(V_{\rm ds})$. The matrix
\begin{equation}\label{eq:gamma}
{\bm \Gamma}_{L,R}(E)=i\left({\bm \Sigma}_{L,R}(E) - {\bm \Sigma}_{L,R}^\dagger(E) \right)
\end{equation}
accounts for the level broadening due to the coupling to the leads.

In addition to self-consistent computation of spin-resolved electron density, which is required both in equilibrium and nonequilibrium, evaluation of Eq.~(\ref{eq:lb}}) requires to compute also the self-consistently developed electric potential profile~\cite{Ke2004} due to the passage of current. The profile enters into MFAH Hamiltonian~(\ref{subeq:HubbardHamiltonian}) through $v_{\bf i}$ term. This ensures gauge invariance of \mbox{\em I-V} characteristics, i.e., its invariance with respect to the shift of electric potential everywhere by a constant.~\cite{Christen1996} The technical issues in converging nonequilibrium charge densities through self-consistent loop are discussed in Appendix~\ref{sec:appendix_a}.

The \mbox{\em I-V} characteristics of two-terminal ZGNR devices with both transparent and $\approx 60$~k$\Omega$ resistive contacts are shown in Fig.~\ref{fig:IVMediumSizeStrip}. In both devices, at around bias voltage $V_t \approx 0.44$ V, current jumps abruptly by an order of magnitude. To understand the origin of the jump, we plot the transmission function $T(E,V_{ds})$ in Fig.~\ref{fig:IVMediumSizeStrip} just before [panels (b) and (e)] and just after [panels (c) and (f)] the discontinuity has occurred. These plots reveal insulating state on the low voltage side $V_{ds} < V_t$, where transmission probability is exponentially suppressed within the gap region. On the other hand, finite transmission probability appears in the metallic state on the high voltage side $V_{ds} > V_t$ of such voltage-driven nonequilibrium phase transition.~\cite{Kroha2007} When the transparency of the contacts is reduced in Figs.~\ref{fig:IVMediumSizeStrip}(e) and ~\ref{fig:IVMediumSizeStrip}(f), the transmission function acquires sharp peaks due to quantum interference effects in the absence of dephasing (e.g., the amplitude of the resonant mode builds up when electron waves leaking from the quasi-bound state in the ZGNR channel cancel the incident waves and enhance the transmitted ones), which is akin to resonant transmission through double barrier structures.~\cite{Haug2007}

The spin density corresponding to point {\bf B} in Fig.~\ref{fig:IVMediumSizeStrip} is plotted in Fig.~\ref{fig:ChargeAndSpinDensity}(a), demonstrating that insulating state for $V_{ds} < V_t$ is magnetically ordered in a similar fashion as in equilibrium. The AF configuration in nonequilibrium shows that a small amount of spin polarization is present in the middle of the ribbon, as is the case of equilibrium AF configuration whose edge states penetrate deeper (when compared to FM configuration) into the bulk.~\cite{Pisani2007}

Further microscopic insight about the charge dynamics of ZGNR driven by finite bias voltage and current flow is revealed by the profiles of electron density in Fig.~\ref{fig:ChargeAndSpinDensity}(b) for the  insulating state and in Fig.~\ref{fig:ChargeAndSpinDensity}(c) for the metallic state, as well as by the electric potential profile for these two states plotted in Fig.~\ref{fig:CoulombPotential}. For example, in the insulating state for both transparent and resistive contacts ZGNR device, the potential profile in Fig.~\ref{fig:CoulombPotential}(a) and ~\ref{fig:CoulombPotential}(b) is linear, as expected for tunneling.  When the band gap collapses the potential profile shows well-defined voltage drops near the contact regions and almost constant behavior within ZGNR channel, as expected for ballistic conductor. The electric potential profiles along ZGNR in Fig.~\ref{fig:CoulombPotential}(a) can be directly related to spatial distribution of charges in Figs.~\ref{fig:ChargeAndSpinDensity}(b) and \ref{fig:ChargeAndSpinDensity}(c). That is, increased charge density around the contact regions in Fig.~\ref{fig:ChargeAndSpinDensity}(c) for nonmagnetic metallic ZGNR is responsible for the voltage drop being confined (solid blue line) mostly around the contacts in Fig.~\ref{fig:CoulombPotential}(a).

Since abrupt current jump at the threshold voltage, as the most distinctive signature of voltage-driven nonequilibrium phase transition between insulating and metallic states of ZGNR, was not observed in recent experiments~\cite{Li2008,Wang2008a} on sub-10-nm-wide ZGNR nanoribbons, we also investigate how the threshold voltage is affected as the length of the ZGNR channel increases. By doubling the inter-contact distance, from \mbox{6.6 nm} in Fig.~\ref{fig:IVMediumSizeStrip} to \mbox{13.2 nm} in Fig.~\ref{fig:IVLongStrip}, we find that threshold voltage increases from \mbox{$\approx 0.44$ V} at point {\bf B} in Fig.~\ref{fig:IVMediumSizeStrip} to at least \mbox{$\approx 0.77$ V} at point {\bf C} in Fig.~\ref{fig:IVLongStrip}. The current jump by an order of magnitude and electric potential profile in this case are still similar to ZGNR two-terminal device in Figs.~\ref{fig:IVMediumSizeStrip}(a) and ~\ref{fig:CoulombPotential} due to the fact that  metallic nonmagnetic ZGNR above the threshold voltage is attached to highly transparent contacts. Nonetheless, the hysteretic behavior at discontinuity in Fig.~\ref{fig:IVMediumSizeStrip}(a), we predict for short ZGNR attached to electrodes via transparent contacts, is almost completely removed in longer devices.

\begin{figure}
\includegraphics[scale=0.5,angle=0]{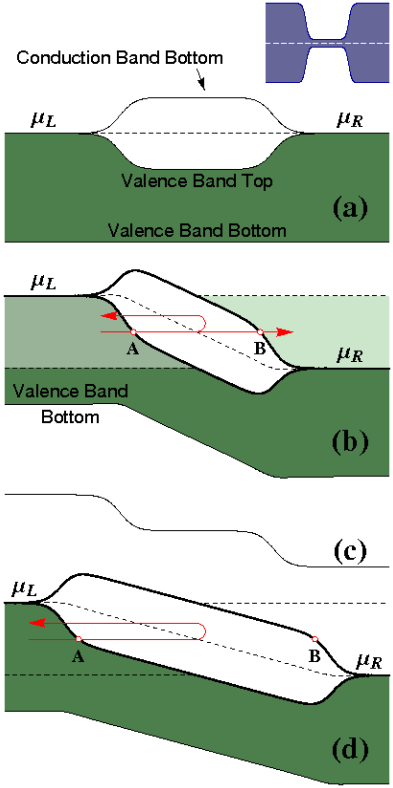}
\caption{(Color online) Schematic explanation of how ZGNR length affects the threshold voltage for nonequilibrium phase transition between its magnetically ordered insulating and nonmagnetic metallic states.  Top left inset depicts the graphene pattern cut out of a single sheet, which could be used to observe band gap collapse and destruction of magnetic ordering in ZGNR by current flow. (a) Schematic plot of the local band gap along the dashed line in the top right inset at zero bias voltage $V_{ds}=0$. (b) Local band gap in the spin-polarized state for short ZGNR under nonequilibrium conditions $V_{ds} \neq 0$. (c) Electrostatic potential profile after the band gap collapse. (d) Local band gap in the spin-polarized state for long ZGNR under nonequilibrium conditions $V_{ds} \neq 0$. Shading is panels (a), (b), and (d) denotes the occupancy of electron states---the lighter the shade, the less is the occupation probability, where dark color corresponds to occupation probability one.}\label{fig:SchematicLengthDependence}
\end{figure}

The decrease of the ZGNR band gap size is not a gradual process, but it is triggered when the bias voltage reaches a specific value. This is particularly transparent in infinite ideal ZGNR (with no voltage drop along ZGNR) of Sec.~\ref{sec:InfiniteZGNR} where the reduction of the band gap and diminishing of edge magnetization density is initiated when the bias voltage window becomes equal to the band gap value. With further increase of the bias voltage, the gap and spin density decay quickly to zero.~\cite{Gunlycke2007} On the other hand, in finite-length ZGNR, the region of negligible transmission function $T(E,V_{ds})$ in Fig.~\ref{fig:IVLongStrip} increases from panel (b) to panel (c) with increasing $V_{ds}$ from point {\bf B} to threshold voltage at around point {\bf C}. This is due to electrostatic potential tilting of the local band structure, so that band gaps in different regions of the ZGNR cover different energy ranges inside the window where we observe $T(E,V_{\rm ds}) \rightarrow 0$ in Figs.~\ref{fig:IVLongStrip}(c) and (d).

Although one can expect that the values of the voltage at which nonequilibrium phase transition takes place will increases with the thickness of the tunnel barrier region introduced by magnetically ordered ZGNR, the values of the threshold voltage (or, more appropriately, a window of voltages taking into account hysteretic behavior) one can expect for realistic two-terminal devices are non-trivial. For example, in abstract infinite ZGNR of Sec.~\ref{sec:InfiniteZGNR} this value is limited~\cite{Gunlycke2007} to \mbox{$\approx 0.4$ V}. On the other hand, in realistic two-terminal devices we find that the threshold voltage increase with increasing length of ZGNR, whose schematic explanation is provided by Fig.~\ref{fig:SchematicLengthDependence} depicting the band energy diagrams for the narrow ZGNR bridging the two contacts. The spin-polarized state becomes unstable when the occupancy of the electron levels in the valence band decreases below, and the conduction band occupancy increases beyond the threshold level.  Under nonequilibrium conditions, the change in occupancy in the short strips becomes possible due to the tunneling through the band gap [Fig.~\ref{fig:SchematicLengthDependence}(b)], which results in the subsequent band gap collapse [Fig.~\ref{fig:SchematicLengthDependence}(c)].  In contrast, the tunneling rate through the band gap in long strips may be too low, and the required change in electron population cannot be achieved for a given $V_{ds}$, as illustrated by \mbox{Fig.~\ref{fig:SchematicLengthDependence}(d)}.

\subsection{Finite-length ZGNR attached to square lattice leads} \label{sec:FiniteZGNRsquare}

\begin{figure}
\includegraphics[scale=0.32,angle=0]{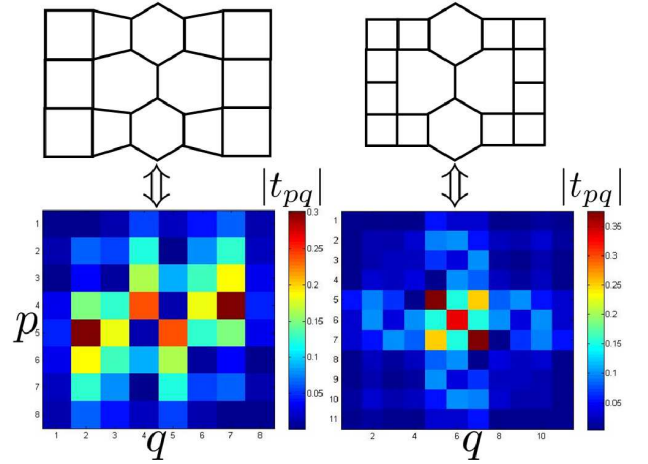}
\caption{(Color online) Color-coded amplitudes of the transmission matrix elements $|t_{pq}(E)|$  connecting transverse propagating modes at $E_F=10^{-6}t$ in the left and right square lattice leads which are ``lattice-unmatched'' (left column) or ``lattice-matched'' (right column) to 8-ZGNR with collapsed band gap. The ZGNR width $N_z=8$ and length $N_z^a=37$ are the same as in the two-terminal device studied in Figs.~\ref{fig:SpinDensityShortSquareLatticeLeads} and \ref{fig:SquareLeads}.}\label{fig:leads}
\end{figure}

Since Sec.~\ref{subsec:FiniteZGNR} suggest that the magnitude of abrupt current jump at the threshold voltage  will depend on the quality of contacts through which the finite-length ZGNR is attached to external circuit, in this section we examine two-terminal devices whose metallic leads, modeled as semi-infinite square lattice wire, are attached laterally (rather than vertically on the top of ZGNR contact region as in Sec.~\ref{subsec:FiniteZGNR}). Several different ways of attaching square lattice leads to the honeycomb lattice of ZGNR have been explored in a variety of recent quantum transport studies.~\cite{Robinson2007,Dragomirova2008a,Sheng2005b} For example, one can attach ``lattice-matched''~\cite{Robinson2007} or ``lattice-unmatched''~\cite{Dragomirova2008a,Sheng2005b} leads illustrated in Fig.~\ref{fig:leads}. The former case is matched in the sense that the lattice constant of such lead is equal to carbon-carbon distance in graphene.~\cite{Robinson2007}

\begin{figure}
\includegraphics[scale=0.6,angle=0]{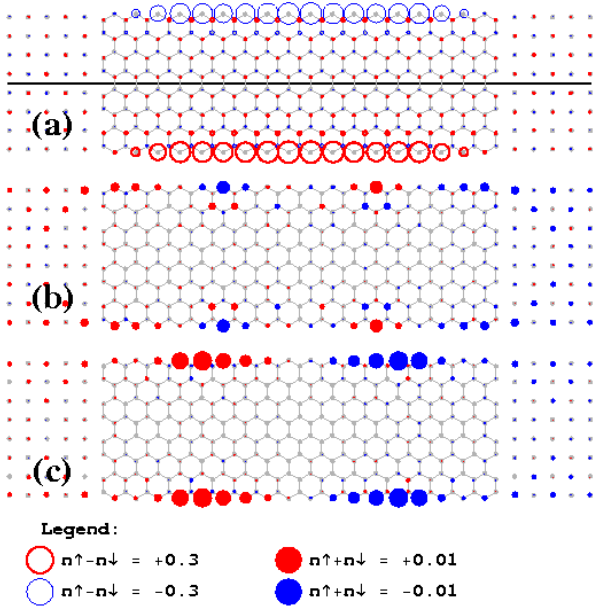}
\caption{(Color online) Spatial profile of spin [panel (a)] and charge [panels (b) and (c)] density within finite-length \mbox{8-ZGNR} (width $\approx 1.5$ nm; length $\approx 4.4$ nm) connected to square lattice leads at the threshold voltage $V_{ds} \approx 0.4$~V: (a) spin density in the state marked by point \textbf{B} in \mbox{Fig.~\ref{fig:SquareLeads}}; (b) charge distribution in the same state as in panel (a); and (c) charge distribution in the non-polarized state marked by point \textbf{C} in \mbox{Fig.~\ref{fig:SquareLeads}}.}\label{fig:SpinDensityShortSquareLatticeLeads}
\end{figure}
\begin{figure}
\includegraphics[scale=0.6,angle=0]{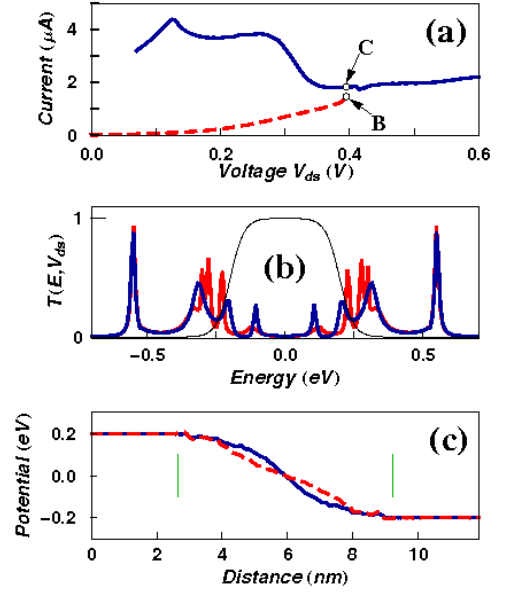}
\caption{(Color online) (a) The \mbox{\em I-V} characteristics, (b) transmission function, and (c) electric potential profile for \mbox{8-ZGNR} (width $\approx 1.5$ nm; length $\approx 4.4$ nm) two-terminal device depicted  in Fig.~\ref{fig:SpinDensityShortSquareLatticeLeads}. The hopping parameter on both square and honeycomb lattice is  $t_{sl}=t=2.7$ eV. In panels (a) and (c), dashed red line denotes AF insulating state of ZGNR, while solid blue line corresponds to metallic state with destroyed spin polarization and collapsed band gap. Panel (b) shows the transmission function in magnetic (red) and nonmagnetic (blue) states of ZGNR marked by points {\bf B} and {\bf C} in panel (a), respectively.  Thin black line illustrates energy window $f(E-eV_{ds}/2)-f(E+eV_{ds}/2)$ over which $T(E,V_{ds})$ is integrated to get the current at the corresponding points {\bf B} and {\bf C}. The potential profile in panel (c) is plotted along the line drawn in the middle of ZGNR in \mbox{Fig.~\ref{fig:SpinDensityShortSquareLatticeLeads}(a)}.  Thin vertical lines in panel (c) mark the boundaries of the extended device ZGNR + portion-of-metallic-leads, shown in \mbox{Fig.~\ref{fig:SpinDensityShortSquareLatticeLeads}}, across which the self-consistent voltage drop is calculated.}\label{fig:SquareLeads}
\end{figure}

When the hopping parameters $t_{sl}=t_c=t$ are selected to be the same in the square lattice region ($t_{sl}$), across the interface ($t_c$), and in graphene ($t=2.7$ eV), the transport across the interface is nominally ballistic, rather than through a tunnel barrier generated by reduced $t_c$ or mismatched $t_{sl}$ and $t$. Nevertheless, the conductance of lead-ZGNR-lead device can be greatly reduced if many propagating modes from metallic leads do not couple well to evanescent modes in GNR, or to evanescent modes plus a single propagating mode at the Fermi energy of ZGNR
with collapsed band gap~\cite{Cresti2007,Zarbo2007} (which are the only available modes to carry transport in narrow ribbons with large gap between $E_F=0$ and the second subband~\cite{Zarbo2007}). Although evanescent modes are effectively enabling doping of GNR by metallic contacts, which is enhanced in short and wide GNRs,~\cite{Tworzydlo2006,Golizadeh-Mojarad2009} the armchair transverse interface of ZGNR coupled to square lattice generates concurrently mixing of transverse propagating modes.~\cite{Robinson2007,Dragomirova2008a} This is illustrated by the non-zero off-diagonal elements of the transmission matrix
\begin{equation}\label{eq:tmatrix}
{\bf t}(E) = \sqrt{{\bm \Gamma}_R(E)} \cdot {\bf G}(E) \cdot \sqrt{{\bm \Gamma}_L(E)}
\end{equation}
in Fig.~\ref{fig:leads} for metallic ZGNR (assuming collapsed band gap) at the Dirac point $E_F=10^{-6}t$, which is equivalent to the presence of disorder at the interface.

Furthermore, the mode mixing is stronger, with larger off-diagonal ${\bf t}$-matrix elements, for ``lattice-unmatched'' leads in Fig.~\ref{fig:leads}. Therefore, the corresponding linear response resistance $R = \displaystyle\frac{h}{e^2} \left(\sum_{pq} |t_{pq}|^2\right)^{-1}$ of the device in Fig.~\ref{fig:SpinDensityShortSquareLatticeLeads} (assuming collapsed band gap) is $R \approx 2.7$~$h/e^2$ (the resistance quantum $h/e^2$ is 25.8 k$\Omega$). On the other hand, the corresponding device with the same \mbox{8-ZGNR} channel length ($N_z^a=37$ or $\approx 4.4$ nm) and ``lattice-matched'' leads in the right column of Fig.~\ref{fig:leads} has $R \approx 2.29$~$h/e^2$. The decay of the resistance with the \mbox{8-ZGNR} channel length in the absence of defects, edge scattering, impurities, and acoustic phonons (all of which were taken into account to extract the mean free path and contact resistance from $R$ vs. $N_z^a$ plot in Ref.~\onlinecite{Wang2008a}) is due to reduced overlap of evanescent modes injected by two metallic electrodes.~\cite{Golizadeh-Mojarad2009} For realistic contacts between various metals and graphene, one would also have to control the alignment of differing energy levels at the interface in order to reduce the effect of the Schottky barrier on the contact resistance.~\cite{Wang2008a,Lee2008}

We choose the setup with ``lattice-unmatched'' leads in Fig.~\ref{fig:SpinDensityShortSquareLatticeLeads} as the device with lower contact transparency, to investigate their effect on the observability of voltage-driven collapse of edge magnetic ordering  and the band gap corresponding to it. The current jump in Fig.~\ref{fig:SquareLeads} is much less pronounced than in Figs.~\ref{fig:IVMediumSizeStrip} and \ref{fig:IVLongStrip}, but it is still observable. This demonstrates that the value of the contact resistance itself, which is generated by different transport mechanisms for devices in Figs.~\ref{fig:ChargeAndSpinDensity} and ~\ref{fig:SpinDensityShortSquareLatticeLeads}, is not the only reason for reducing the current jump at the threshold voltage. Moreover, despite poor contact between metallic leads and ZGNR, we still find hysteretic behavior in Fig.~\ref{fig:SquareLeads}, which (assuming that it is not an artifact of self-consistent convergence algorithm for nonequilibrium electron densities, see Appendix~\ref{sec:appendix_a}) could also be used to confirm the band gap collapse in the nonequilibrium state of ZGNR.

One can also compare the charge density redistribution in Fig.~\ref{fig:SpinDensityShortSquareLatticeLeads} and electric potential profile in Fig.~\ref{fig:SquareLeads} to those in Sec.~\ref{subsec:FiniteZGNR} for a two-terminal ZGNR device with different type of contacts. To ensure that electric potential approaches the constant values in the bulk of the
electrodes, a portion of the square lattice leads are attached to ZGNR to form an ``extended device''~\cite{Ke2004} shown in Fig.~\ref{fig:SpinDensityShortSquareLatticeLeads} for which
self-consistent calculations are performed. Thus, the charge transfer and potential disturbance caused by ZGNR are screened off outside the extended device region [voltage drop within the
extended device region is enclosed by two vertical lines in Fig.~\ref{fig:SquareLeads}(c)], so that potential at the edges of the extended device region matches to constant potential along the ideal semi-infinite electrodes.

\section{Conclusions} \label{sec:conclusion}

In summary, we predict voltage-driven nonequilibrium phase transition between magnetically ordered (Slater) insulating state and nonmagnetic metallic state in finite-length zigzag graphene nanoribbons. The ZGNR is attached to two metallic electrodes, where finite bias voltage brings such two-terminal device into a nonequilibrium steady-state with current flowing through it. The high density of states at the Fermi level, due to special topology of zigzag edges, results in instability when Coulomb interactions are taken into account, which is resolved through spin polarization around the edges in equilibrium. The spin-polarized state survives in nonequilibrium when the current flowing through ZGNR is small. However, at finite threshold voltage, the edge magnetic ordering is destroyed together with the band gap determined by the staggered potential of the magnetization density profile.

The abrupt jump of the current at the threshold voltage, as one of our principal predictions, has unique and experimentally observable features: ({\em i}) current can increase by an order of magnitude when the bias voltage is tuned across the threshold voltage; ({\em ii}) a hysteretic behavior in the \mbox{\em I-V} curve can occur around the window of threshold voltages; ({\em iii}) the value of the threshold voltage increases with increasing nanoribbon length; ({\em iv}) the magnitude of the current discontinuity is reduced for poorly transparent contacts or metallic electrodes whose transverse propagating modes do not match well to ZGNR modes (propagating or evanescent) that can carry current.

While these unique features can be tested with devices amenable to presently nanofabrication technology, they have not been observed in recent experiments~\cite{Li2008,Wang2008a} on sub-10-nm-wide graphene nanoribbon-based field-effect transistor devices which have displayed a sizable band gap for bias voltages up to \mbox{1 V}. Thus, we delineate two prerequisites for observing the collapse of the band gap and underlying magnetic ordering through the measurement of \mbox{\em I-V}  characteristics of ZGNR two-terminal devices:
\begin{enumerate}
\item The contact resistance between ZGNR and metallic electrodes should be kept low.  This does not imply that the contacts must be perfectly transparent.  Our simulation results indicate that the \mbox{\em I-V} curve signatures of the band-gap collapse should be observable even when the contact transparency constitutes 20\% from the ideal value, which corresponds to experimentally measured contact resistance of $\simeq 60$~k$\Omega$ in GNRFETs with top-deposited Pd electrodes.~\cite{Wang2008a}

\item The threshold voltage and the corresponding threshold current required to collapse the  band gap of sub-10-nm wide ZGNRs increase with increasing of nanoribbon length.  The threshold source-drain voltage for sub-10-nm-wide ZGNR which is $\approx 6.6$~nm long is $\approx 0.44$~V, and for ZGNR which is twice as long $\approx 13.2$~nm the threshold voltage $V_t$ increases to $\approx 0.8$~V.  At the same time, the shortest ribbon for which the experimental \mbox{\em I-V} curve measurements were performed~\cite{Wang2008a} was \mbox{$\approx 110$ nm} long and the applied source-drain voltage did not exceed $1$~V.  This suggests that the threshold criteria has not been met.  The decrease of the ZGNR length down to \mbox{10--20~nm} range could result in  experimental observation of predicted current-flow-induced transition between spin-polarized and non-polarized ZGNR states.
\end{enumerate}

Thus, fabricating proposed devices---short graphene nanoribbon with atomically ultrasmooth zigzag edges~\cite{Li2008} sandwiched between as transparent metallic contacts as possible---can be used to detect the presence of unusual {\em s-p} magnetism of carbon atoms along zigzag edges in unambiguous fashion and with {\em all-electrical} setup. If the predicted band gap collapse can be induced in such devices, the \mbox{\em I-V} curve measurements will also probe aspects of spin dynamics in ZGNR. For example, if the current is turned off after ZGNR has been transformed from spin-polarized semiconducting to non-polarized semimetallic state, the spin-polarized state is expected to be restored with some time delay $\tau_{\rm off}$.  That can be explained as follows: the energy gain associated with small increment of the spin-polarized density at the zigzag edges is proportional to the spin density already accumulated at this edge.  Because the spin density in non-polarized state is zero, no first order driving force is present to transform the system from non-polarized to spin-polarized state.  The delay $\tau_{\rm off}$ can be measured as the time needed for \mbox{\em I-V} curve to change its character from conductive to highly resistive state.  The time delay $\tau_{\rm on}$ in the onset of the band-gap collapse after current is turned on is expected to be much smaller than $\tau_{\rm off}$.  However, if $\tau_{\rm on}$ can be measured, information on electron velocity in the edge states, spin ordering, and the dependence of spin ordering on temperature could be deduced, in principle.

\begin{acknowledgments}
We thank R. L. Dragomirova for illuminating discussions. Financial support from NSF Grant No. ECCS 0725566 is gratefully acknowledged.
\end{acknowledgments}

\appendix

\section{Self-consistent algorithm for nonequilibrium electron density in ZGNR}\label{sec:appendix_a}

We compute the nonequilibrium solution using the voltage step $\Delta V=0.01$~V.  For the magnetically ordered ZGNR, the solution for $V_{ds}^{(i)}$ uses the solution for $V_{ds}^{(i-1)}$ as the starting point.  Calculations for the spin-polarized case start with \mbox{$V_{ds}^{(0)}=0$} and proceed to the point $V_{t}$ beyond which the convergence towards the self-consistent solution cannot be obtained.   Even though the inability to obtain convergence does not constitute a proof of the solution non-existence, we assume that $V_{t}$ reached in this fashion is the threshold voltage at which spin polarization is destroyed.  The Newton-Raphson algorithm described in Sec.~\ref{sec:NewtonRaphson} guarantees the convergence towards the self-consistent solution, provided that sufficiently small portion of $\delta {\bf n}_{\rm in}$ obtained from Eq.~(\ref{eq:SCConditionLinearSystem}) is used to augment ${\bf n}_{\rm in}$ of the previous iteration.  This self-consistent solution is not guaranteed to correspond to the lowest energy and depends on the initial electron density.  Our algorithm monitors the convergence and adjusts the step along $\delta {\bf n}_{\rm in}$ vector to ensure that the ${\bf n}_{\rm in}$ is changed in a such way that the difference between ${\bf n}_{\rm in}$ and ${\bf n}_{\rm out}$ decreases during each iteration.

When the step $\alpha$ becomes too small, i.e., the input density correction $\alpha \times \delta {\bf n}_{\rm in}$ corresponds to the maximum potential variation of less than \mbox{1~meV} that is still not small enough to make the difference between ${\bf n}_{\rm in}$ and ${\bf n}_{\rm out}$ in a current iteration less than in a previous one, the calculation stops.  Even if the self-consistent solution can be obtained by further decreasing the step $\alpha$, it will be highly unstable with respect to the external perturbations of the order of \mbox{1~meV}.  In a usual experimental environment this is equivalent to non-existence of spin-polarized solution.

The non-polarized solution is obtained for sufficiently high applied voltage, such as $V_{ds}=1.0$~V, to ensure the solution stability.  Then, the solutions for the lower voltages are computed using the higher voltage solution as the starting points.  The calculations stop when the same criteria for the threshold voltage as in the case of spin-polarized solution is met.  This  can lead to well-pronounced hysteretic behavior of \mbox{\em I-V} curves, depending on the contacts, as illustrated by Figs. \ref{fig:IVMediumSizeStrip}(a), \ref{fig:IVLongStrip}(a), and \ref{fig:SquareLeads}(a).




\end{document}